\documentclass[reprint,aps,prd,superscriptaddress]{revtex4-1}
\usepackage[utf8]{inputenc}
\usepackage{amsmath}
\usepackage{natbib}
\usepackage{amsfonts}
\usepackage{hyperref}
\usepackage{cleveref}
\usepackage{amssymb}
\usepackage{graphicx}
\usepackage[usenames,dvipsnames]{color}
\usepackage{mathrsfs}

\newcommand{\abs}[1]{\left\vert#1\right\vert}
\newcommand{\set}[1]{\left\{#1\right\}}
\newcommand{\Real}{\mathbb R}
\newcommand{\eps}{\varepsilon}

\newcommand{\ret}{\textrm{ret}}

\def\actaa{\ref@jnl{Acta Astron.}}      
\def\apj{\textrm{ApJ}}                 
\def\apjl{\textrm{ApJ}}                
\def\aap{\textrm{A\&A}}                

\def\mnras{\textrm{MNRAS}}             
\def\prd{\textrm{Phys.~Rev.~D}}        
\def\prl{\textrm{Phys.~Rev.~Lett.}}    
\def\physrep{\textrm{Phys.~Rep.}}   

\begin{document}

\title{The Probability Distribution of Astrophysical Gravitational-Wave Background Fluctuations}

\author{Yonadav Barry Ginat}
\email{ginat@campus.technion.ac.il}
\affiliation{Physics department, Technion, Haifa 3200003, Israel}
\author{Vincent Desjacques}
\email{dvince@physics.technion.ac.il}
\affiliation{Physics department, Technion, Haifa 3200003, Israel}
\affiliation{Asher Space Science Institute, Technion, Haifa 3200003, Israel}
\author{Robert Reischke}
\affiliation{Physics department, Technion, Haifa 3200003, Israel}
\affiliation{Department of Natural Sciences, The Open University of Israel, 1 University Road, P.O. Box 808, Ra'anana 4353701, Israel}
\author{Hagai B. Perets}
\affiliation{Physics department, Technion, Haifa 3200003, Israel}
\affiliation{Asher Space Science Institute, Technion, Haifa 3200003, Israel}
\affiliation{Theoretical Astrophysics 130-33, California Institute of Technology, Pasadena,
CA 91125}

\begin{abstract}
  The coalescence of compact binary stars is expected to produce a stochastic background of gravitational waves (GW) observable with future GW detectors. Such backgrounds are usually characterized by their power spectrum as a function of frequency. Here, we present a method to calculate the full 1-point distribution of strain fluctuations. We focus on time series data, but our approach generalizes to the frequency domain. We illustrate how this probability distribution can be evaluated numerically. In addition, we derive accurate analytical asymptotic expressions for the large strain tail, which demonstrate that it is dominated by the nearest source. As an application, we calculate the distribution of strain fluctuations for the astrophysical GW background produced by binary mergers of compact stars in the Universe, and the distribution of the observed confusion background obtained upon subtracting bright, resolved sources from the signal.
  We quantify the extent to which they deviate from a Gaussian distribution. Our approach could be useful for the spectral shape reconstruction of stochastic GW backgrounds.
\end{abstract}

\maketitle

\section{Introduction}
\label{sec:introduction}

The direct discovery of gravitational waves (GWs) from binary black hole mergers in 2015 \cite{LIGOVirgo2016} sparked a new interest in gravitational waves, which constitute a new window to the Universe. Due to the relative weakness of gravity, the amplitudes of gravitational waves are rather small, and many GW-emitting processes pass under our noses undetected. Their cumulative effect amounts to a gravitational-wave background that bathes the detectors (for a recent review see \citep{2018CQGra..35p3001C}) and may, when investigated, reveal details of its physical origin (see for instance \citep{KosenkoPostnov2000,FarmerPhinney2003,FaltaFisher2011,Thrane2013,Callisteretal2016,Cusinetal2017,Brito:2017wnc,Jenkinsetal2018,Barausse:2018vdb,Christensen2019,Conneelyetal2019} for astrophysical backgrounds). This background shares a lot of properties with other cosmic backgrounds, such as the cosmic microwave background (e.g. their stochastic nature), but is also unique, in ways we will explore here.

That the stochastic gravitation-wave background (SGWB) from cosmological sources (such the primordial GWs produced during inflation) is a Gaussian random field is well-known \cite{Maggiore2}. The situation is less clear for backgrounds of astrophysical origin, although, \emph{prima facie}, the Central Limit Theorem suggests that the distribution of observed strain should converge towards a Gaussian when the number of sources, $N$, tends to infinity. As opposed to primordial GWs, which are generated by inherently random quantum fluctuations during inflation, astrophysical sources are purely deterministic, in-so-far-as their position, separation, masses, \emph{etc.} specify the wave-forms fully. Hence, while in the former case the wave amplitudes themselves are random, in the latter -- which is our main concern in this paper -- the only randomness is in the spatio-temporal location of the source on the past light cone of the observer (as well as its position in the relevant parameter space). Therefore, one expects that a superposition of signals from such sources need not be a Gaussian random field, unless $N$ becomes very large.

So far, most of the literature on SGWBs has focused on the quantity
\begin{equation}
  \Omega_{\textrm{gw}}(f) = \frac{1}{\rho_{\textrm{crit}}}\frac{\mathrm{d}\rho_\textrm{gw}}{\mathrm{d}\ln f},
\end{equation}
where
\begin{equation}
  \rho_\textrm{gw} = \frac{c^2}{32\pi G}\langle \dot{h}_{ij} \dot{h}^{ij}\rangle
\end{equation}
is the energy density of the observed GW strain $h$ at the (observer-frame) frequency $f$. The density parameter $\Omega_\textrm{gw}$, integrated over all frequencies, is then the \emph{mean} SGWB energy density signature. In this paper, we aim at providing a tractable approach for determining the full (1-point) probability distribution function (PDF) of $h$ around this mean energy. For simplicity, we will focus on strain time series $h(t)$ analogous to those simulated in e.g.\cite{Coward:2002ba,Regimbau:2011rp,LIGOVIRGO2018a}, but we emphasize that our approach is capable of resolving the frequency dependence of the full distribution function (see Sec.~\S\ref{sec:discussion} for a brief discussion of this point).

In order to carry out this task, we exploit the fact that GWs propagate over distances much larger than the typical clustering length $r_\xi$ of the sources. Therefore, Poisson clustering should provide a good approximation when the bulk of the sources lie at distances $r\gg r_\xi$ from the detector. This simplification enables us to calculate the PDF of the observed strain from the knowledge of the characteristic functions of individual sources solely. In particular, we are interested in by the asymptotic behavior of the strain distribution in the large strain limit, the impact of interferences and the validity of the Gaussian approximation. Furthermore, as a demonstration of the applicability of our methods, we will apply them to the characterization of the PDF of the SGWB produced by binary mergers of compact stars.

The paper is organized as follows: we lay down our assumptions and spell out our approach in Sec.~\S\ref{sec:general}. We then move on to exemplify our method with a toy model in Sec.~\S\ref{sec:simple case}, which we subsequently expand in Sec.~\S\ref{sec:physical case} in order to estimate the probability distribution of the SGWB produced by binary mergers of compact stars in the Universe. In Sec. \S 5 we consider the removal of bright sources form the background estimation. We discuss our results, along with a number of possible extensions, in Sec.~\S\ref{sec:discussion} before concluding in Sec.~\S\ref{sec:conclusions}.
Throughout, we assume that space-time is described by a flat Friedman-Lema\^{i}tre-Robertson-Walker (FLRW) metric, with cosmological parameters $\Omega_m = 0.32, \Omega_\textrm{rad} = 9.187\times 10^{-5}, \Omega_\Lambda = 1 - \Omega_m - \Omega_\textrm{rad}, h = 0.674$ \cite{Planck2018}.

\section{General Considerations}
\label{sec:general}

Suppose that a gravitational-wave detector, located at the origin of the coordinate system, receives signals from $N$ sources distributed uniformly within a sphere of (co-moving) radius $R$. The sources radiate signals -- each of which is characterized by its own relevant physical processes -- which travel to the detector, where they interfere to produce a total signal \cite{Maggiore}
\begin{align}
    s(t) = h(t) + n(t), ~~ & \mbox{with} & ~~
  h(t) \equiv \sum_{\textrm{source}~i} h_i(t).
\end{align}
Here, $n(t)$ is the detector noise (which is uncorrelated with $h(t)$), and $h_i(t)$ is a linear combination of the two different polarizations $h_+$ and $h_\times$ of the waves \footnote{We assume that the total observational time is much larger than the periods of the waves received by the detector.}.
Presupposing that the detector noise can be mitigated (using advanced interferometers like {\small LISA} and/or cross-correlation among multiple detectors), the strain time series $h(t)$ (sampled into small time intervals) -- which contains the largest amount of information about the GW background (down to the residual noise level) -- could be extracted from the data (possibly sampled into small time intervals).

Our goal is to find a way to calculate the 1-point probability distribution (PDF) $P(h)$, which gives the probability of measuring a strain $[h,h+dh]$ at the detector, and to comprehend its basic properties.

For this purpose, we make the following assumptions:
\begin{enumerate}
  \item Spacetime is described by a flat FLRW metric in co-moving coordinates $(t,\mathbf{r})$, with cosmological parameters as specified in Sec.~\S\ref{sec:introduction}. Henceforth, $t$ will stand for the cosmic time while $\eta$ will denote the conformal time. The sources and the detector are idealized co-moving frames.
  \item The GW sources are independent of each other, and are all instances of the same type of sources (say, compact binaries), but may have varying physical parameters (say, chirp masses).
  \item The sources are distributed homogeneously in co-moving space according to an isotropic Poisson process. Therefore, the number $N$ of sources within a sphere of co-moving radius $R$ is Poisson distributed with a mean count $\lambda(R)$: $N \sim \textrm{Pois}(\lambda(R))$.
  \item Each source has a probability density $R_*(t)$ of turning on at a given cosmic time $t$.
  \item A source which turned on at $t=t_*$ at a co-moving distance $r=|\mathbf{r}|$ from the detector produces a strain at $r=0$ and time $t_0>t_*$, which is given by a known function $g$:
      \begin{equation}\label{eqn:gw propagation}
        h(t) = \frac{d}{d_L(r)} e^{-r/r_0}g\left(\int_{0}^{\eta_{0,\ret}}\mathrm{d}\eta\,a(\eta),\varphi\right),
      \end{equation}
      $\eta_{0,\ret} = \eta_0 - r/c$ is the retarded conformal time, $\eta_0$ is the conformal time measured by the detector (i.e. the conformal age of the universe today), $d$ is an arbitrary distance chosen to normalize the strain (to make $g$ dimensionless), and $\varphi$ is a random phase uniformly distributed in the interval $[0,2\pi]$.
      We include an exponential decay with characteristic length scale $r_0$ for generality.
      This could represent a physical damping caused by an anisotropic dark matter energy-momentum tensor \cite{FlaugerWeinberg2018,FlaugerWeinberg2019} for instance.
  \item The wavelength of the GWs is considerably smaller than the radius of curvature of the Universe. As a result, the polarizations decouple and propagate along null-geodesics \cite[\S 4.1.4]{Maggiore}.
\end{enumerate}

Let us comment briefly on some of these assumptions: in general, the properties of the GW signal $g$ depends on a set of model parameters (which include, e.g., the chirp mass of a binary star), $\xi = (\xi_1,\xi_2,\ldots)$. As mentioned earlier, one expects that the distant astrophysical sources be clustered spatially (a phenomenon which also leads to anisotropies on large angular scales, as recently discussed in the literature by, e.g., \citep{KosenkoPostnov2000,Cusinetal2017,Jenkinsetal2018,Cusinetal2018,Jenkinsetal2019a,Jenkinsetal2019b,Bertacca:2019fnt}). However, since the GWs propagate over large distances without being attenuated significantly (any viable $r_0$ is much greater than $r_\xi$), it is reasonable to assume that the sources follow Poisson clustering.
When $r_0$ is large but finite, it is convenient to work with the mean number $N_0$ of sources inside a sphere of radius $r_0$. As the source counts are also homogeneous along the radial direction, one has $N_0 = \lambda(R)(r_0/R)^3$.

In practice, $h$ is a superposition of independent contributions, so it is natural to calculate the Fourier transform of its PDF, $P(h)$, which decomposes into a product of the characteristic functions of the waves emitted by the individual sources \cite{Zhuo1992,fardal/shull:1993,MeiksinWhite2003}. For a spatial Poisson process, the probability to find a \emph{single} source at position $\mathbf{r}$ with parameters $\xi$ and phase $\varphi$ is
\begin{equation}
\label{eq:pdfindividualsource}
    p(\mathbf{r},\xi,\varphi)=\frac{4\pi r^2 \mathrm{d}r}{\frac{4}{3}\pi R^3}\cdot \phi(\xi) \mathrm{d}\xi\cdot\frac{\mathrm{d}\varphi}{2\pi} \;.
\end{equation}
Here and henceforth, $\phi(\xi)$ denotes the measure on the source parameter space \footnote{Throughout the paper, we use $\phi(\cdot)$ to denote the probability measure on any subset of the model parameters $\xi$.}

Let $\psi(q)$ be the corresponding characteristic function -- the PDF's Fourier transform --
\begin{equation}\label{eqn:volume average}
  \psi(q) = \frac{3}{4\pi R^3}\int_{0}^{R} \mathrm{d}r\,r^2\int_{S^2}d\Omega\int \mathrm{d}\xi\, \tilde{p}(q;\xi;\mathbf{r})\;,
\end{equation}
where
\begin{equation}\label{eqn:n_f}
  \tilde{p}(q;\xi;\mathbf{r}) = \int_{0}^{2\pi}\frac{\mathrm{d}\varphi}{2\pi}\exp\left[\frac{\mathrm{i}qde^{-\frac{r}{r_0}}}{d_L(r)} g\left(\int_{0}^{\eta_{0,\ret}}\mathrm{d}\eta\,a(\eta), \varphi\right)\right]
\end{equation}
is the phase-average of $\exp(\mathrm{i} q h(t))$ for a single source.

Since the sources are identical, they share the same characteristic function $\psi(q)$, whence the characteristic function $\Xi(q)$ for Poisson distributed sources in a sphere of radius $R$ is the Poisson mixture
\begin{align}
\Xi(q) &= \sum_{k=0}^\infty \left(\frac{\lambda^k}{k!}e^{-\lambda}\right)\cdot \psi(q)^k \nonumber \\
&= e^{\lambda \psi(q)-\lambda}.
\end{align}
(One could start the sum at $k=1$ and end up with an irrelevant additive constant term which can be dropped.) Consequently, the PDF of the observed strain produced by Poisson distributed sources is given by the inverse Fourier transform
\begin{equation}
\label{eqn:stochastic gw probability density}
  P\left(\sum h_k = h\right) = \frac{1}{2\pi}\int_{-\infty}^{\infty}\mathrm{d}q\,e^{-\mathrm{i}qh}e^{\lambda\psi(q) - \lambda}\;.
\end{equation}
The main question is how to obtain a meaningful expression for $\psi(q)$. As we will show below, the asymptotic properties of $P(h)$ can be gleaned from equation \eqref{eqn:stochastic gw probability density} without explicit knowledge of $\psi(q)$, as they rely only on universal properties of the physical system. For example, provided that each source has an equal probability of emitting $+h_k$ as $-h_k$, $\psi(q)$ is an even function of $q$ and, therefore, $P(h)$ is also an even function of $h$. This implies that $P'(h=0) = 0$, i.e. $P(h)$ flattens for small values of $|h|$.

An analytical derivation of the form of $P(h)$ at large $h$ is somewhat more challenging, and is expounded below. The results are reminiscent of cosmological results obtained with the theory of large deviations (see \citep{touchette:2009} for a review), when applied in the context of the large scale structure (LSS; see, for instance, \citep{balian/schaeffer:1989,bernardeau/schaeffer:1999,valageas:2002,valageas/munshi:2004,bernardeau/etal:2014a,uhlemann/etal:2017,Barthelemy:2019ciu}).
At this point, it is worthwhile to draw an analogy between the physical system considered here and random walks. Since the observed strain is a superposition of independent but identical waves, its time evolution is analogous to a random walk in the complex plane: the position of the walker after $n$ steps is of the form $\sum_{k=1}^{n}A_k\exp(\mathrm{i}\varphi_k)$, which is precisely a sum of $n$ random waves, with $\varphi_k$ representing the phase of the $k$-th wave as it reaches the detector (see \cite{BarryHughes1995,Hui:2012yp}). Here, however, the amplitudes of the waves have a unique property -- they all obey the law of gravity (they decay as $1/r$), but \emph{ipso facto} one will be able to deduce general properties of the PDF.

To exemplify our method and illustrate the key properties of $P(h)$, it is instructive to consider a simplified scenario first. This is the focus of Sec.~\S\ref{sec:simple case}. A more realistic calculation will be carried out in Sec.~\S\ref{sec:discussion}.

\section{A simplified Model}
\label{sec:simple case}

In order to understand the features of $P(h)$, we consider a simple test case in which the GWs emitted are all described by a pure sine wave with a constant frequency $\omega$, i.e. $g(t) = A\cos(\omega t + \varphi)$. Furthermore, we ignore any cosmological effects, but assume there is a finite attenuation length $r_0$. Picking up $d\equiv r_0$, the strain produced by a single source at the detector is
\begin{equation}
\label{eq:gth}
  g(t,\varphi) = \frac{Ar_0e^{-r/r_0}}{r}\cos(\omega t +\varphi) \;.
\end{equation}
The source parameters are $\xi=(A,\omega)$ in this case. We leave the distributions of $A$ and $\omega$, $\phi(A)$ and $\phi(\omega)$, undetermined. While this simplified model serves its purpose of demonstrating the salient points of the mathematical technique used in this paper, it does not, however, constitute a simplified version of the physical case discussed below in \S \ref{sec:physical case}.

\subsection{The Characteristic Function}

The phase-average of $\exp(\mathrm{i} q h(t))$, needed for the computation of the single source characteristic function, is
\begin{align}
  \tilde{p}(q;A,\omega;r) &
  = \int_{0}^{2\pi}\frac{\mathrm{d}\varphi}{2\pi}\exp\left[\mathrm{i}q\frac{Ar_0e^{-r/r_0}}{r}\cos(\omega t +\varphi)\right] \nonumber \\ &
  = J_0\!\left(q\frac{Ar_0e^{-r/r_0}}{r}\right)
  \label{eq:tildepsimple},
\end{align}
where $J_0$ is the zeroth (cylindrical) Bessel function.

The characteristic function $\psi(q)$ is obtained upon a further average over the volume of a sphere of radius $R$ (corresponding to the maximum distance at which sources can form, regardless of the detector sensitivity), the distribution of the amplitude $A$ and, in principle, the frequency $\omega$ (although the latter is immaterial in the 1-point PDF as $\tilde{p}$ is independent of it). Explicitly,
\begin{equation}
  \psi(q) = \int \mathrm{d}\omega\, \phi(\omega) \int \mathrm{d}A\, \phi(A) \frac{3}{R^3}\int_{0}^{R}\mathrm{d}r\, r^2 \tilde{p}(q;A,\omega;r);
\end{equation}
or, upon substituting the variables $\tau \equiv r/r_0$, $s\equiv q h_c$ and $b\equiv A/h_c$,
\begin{equation}
\label{eqn:psi simple case}
 \psi(s) = \frac{3N_0}{\lambda}\int \mathrm{d}b\, \phi(b) \int_{0}^{R/r_0}\mathrm{d}\tau\, \tau^2 J_0\!\left(s\frac{b e^{-\tau}}{\tau}\right).
\end{equation}
The normalization strain $h_c$ is defined through
\begin{align}
  \lambda h_c^2 &\equiv 3N_0\int \mathrm{d}A\, \phi(A)\int_{0}^{R/r_0}\mathrm{d}\tau\, \tau^2\cdot \frac{1}{2}\left(\frac{Ae^{-\tau}}{\tau}\right)^2 \nonumber \\
  &= \frac{3}{4}N_0\langle A^2\rangle(1-e^{-2R/r_0})
  \nonumber \\
  &\approx \frac{3}{4}N_0\langle A^2\rangle \;.
\end{align}
The last approximation is valid only if $R\gg r_0$. A multiplicative factor of $\lambda$ is taken out of $h_c^2$, so that the latter is defined for a mean count of unity. In fact, $\lambda h_c^2\equiv \langle h^2\rangle$ is exactly the variance of strain fluctuations, as is readily visible from computing the second moment of $P(h)$ from the general relation
\begin{equation}
    \langle h^k\rangle = (-\mathrm{i})^k\frac{\mathrm{d}^k}{\mathrm{d}q^k}\Big[e^{\lambda\psi(q)-\lambda}\Big]\bigg\lvert_{q=0} \;.
\end{equation}
The transformation $q\mapsto s$ implies that $h$ is now measured in units of $h_c$. All one has to do to obtain a PDF for $h$ itself is to divide by $h_c$, where needed.

It is evident from equation \eqref{eqn:psi simple case} that the quantity we are interested in, $\lambda\psi(s) -\lambda$, depends linearly on $N_0$. Therefore, it is easier to define $N_0 G(s) = \lambda\psi(s) - \lambda$, so that $G(s)$ is independent of $N_0$. Then
\begin{equation}\label{eqn:G simple case}
  G(s) = 3\int \mathrm{d}b\, \phi(b) \int_{0}^{R/r_0}\mathrm{d}\tau\,  \tau^2 \left[J_0\left(s\frac{b e^{-\tau}}{\tau}\right) - 1\right]\;.
\end{equation}
Note that $G(s)$ is nothing but the cumulant generating function for $N_0=1$. It satisfies $G(0)=G'(0)=0$, while $G''(0)<0$, it reaches its global maximum at $s=0$.

\subsection{The shape of the distribution}

\begin{figure}
  \centering
  \includegraphics[width=0.45\textwidth]{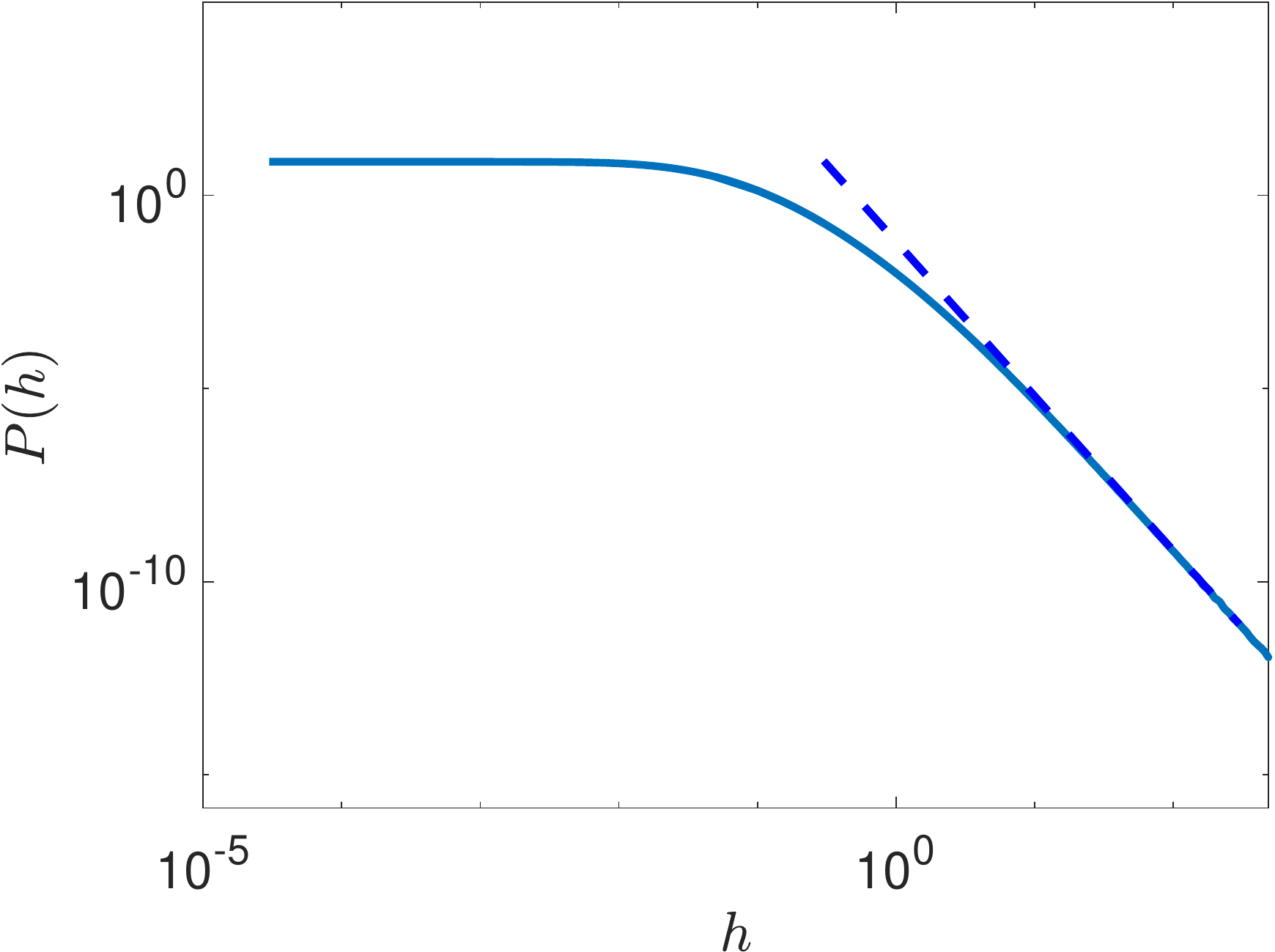}
  \includegraphics[width=0.45\textwidth]{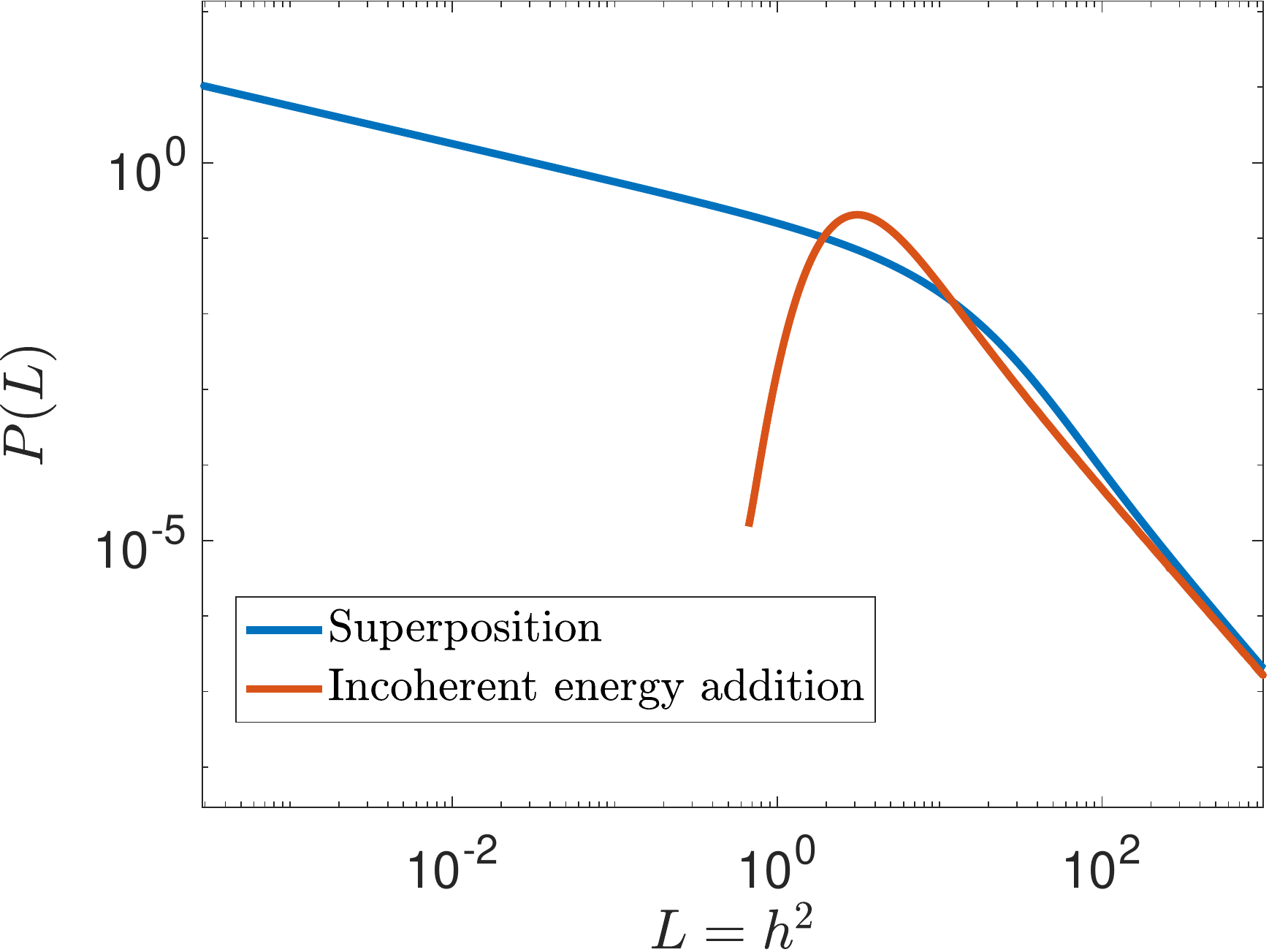}
  \caption{Top: Numerical evaluation of the Fourier transform of $\exp(\lambda\psi - \lambda)$ (solid) along with the leading, large-$h$ asymptotic result (\ref{eqn:h^-4 power law}) for the illustrative model considered in \S \ref{sec:simple case}. Bottom: Comparison between the probability distribution for $h^2$ obtained via equation \eqref{eqn:stochastic gw probability density}, and that for $L$ obtained by summing up the \emph{intensities} of individual source (see text).
  $N_0$ is set to $0.1$ for the top panel, and $10$ for the bottom panel. This leads to $\langle L\rangle = \langle h^2\rangle = 7.5$ in the right panel.
  A distribution $\phi(A)$ peaked sharply at $A=1$ is assumed throughout.}
  \label{fig:example of a special case calculation}
\end{figure}

The large-$h$ asymptotic behavior of $P(h)$ can be obtained for any $N_0$, \emph{vide infra} and appendix \ref{appendix:Mellin transform}.
Suppose first that $N_0$ is small. Taking $h>0$ (without loss of generality since $P(h)$ is even in $h$), one may expand the characteristic function $e^{N_0G(s)} \approx 1+ N_0G(s)$. Ignoring the average over $b$ for the moment, one obtains
\begin{align}
  P(h) & \sim \frac{1}{2\pi}\int_{-\infty}^{\infty} \mathrm{d}s\, e^{-\mathrm{i}sh} (1 + N_0G(s)) \\ &
  = \frac{3N_0}{2\pi}\int_{0}^{R/r_0}\mathrm{d}\tau\,\tau^2\int_{-\infty}^{\infty}\mathrm{d}s\,J_0\!\left(\frac{b s e^{-\tau}}{\tau}\right) e^{-\mathrm{i}sh}\nonumber \;.
\end{align}
The Fourier transform may now be performed easily, \emph{viz.}
\begin{equation}\label{eqn:equation18}
  P(h) \sim \frac{3N_0}{\pi}\int_{0}^{R/r_0}\mathrm{d}\tau \begin{cases}
                                                    \frac{\tau^2}{\sqrt{b^2e^{-2\tau}/\tau^2 - h^2}}, & \mbox{if } \abs{h} < \frac{be^{-\tau}}{\tau} \\
                                                    0, & \mbox{otherwise}.
                                                  \end{cases}
\end{equation}
If $u = \tau e^\tau$, then, provided that $b/\abs{h} \leq R/r_0e^{R/r_0}$ (which ought to be the case for sufficiently large $\abs{h}$),
\begin{equation}
  P(h) \sim \frac{3N_0}{\pi}\int_{0}^{b/\abs{h}}\frac{\mathrm{d}u\,W^3(u)}{b(1+W(u))\sqrt{1-h^2u^2/b^2}},
\end{equation}
where $W(u)$ is Lambert's $W$ function.
If $\abs{h} \gg b$, then the integration domain contains solely small values of $u$, for which $W^3(u) \approx u^3$, whence, to leading order,
\begin{equation}
  P(h) \sim \frac{3N_0}{\pi}\frac{b^3}{h^4}\int_{0}^{1}\frac{\mathrm{d}x\,x^3}{\sqrt{1-x^2}},
\end{equation}
where $x = hu/b$. Upon performing the final integral, the first order asymptotic expansion of $P(h)$ turns out to be
\begin{equation}\label{eqn:h^-4 power law}
  P(h) \sim \frac{2N_0\langle b^3\rangle}{\pi h^4}.
\end{equation}
This meshes well with the numerical evaluation of $P(h)$ accurately in the large-$h$ limit, as can be seen from the top panel of figure \ref{fig:example of a special case calculation}.

We contend that this asymptotic relation still holds even when $N_0$ is not small. We give a simple argument for this here, and present a full derivation in appendix \ref{appendix:Mellin transform}. Away from the origin, $G(s)$ is a smooth ($\mathcal{C}^{\infty}$) function and, consequently, its Fourier transform decays faster than any power-law. Indeed, if we divide up the real line into three intervals: $I_0 = (-\delta,\delta), I_- = (-\infty,-\delta + \eps), I_+ = (\delta - \eps,+\infty)$, where $0 < \eps < \delta \ll 1$, and take a partition of unity $\set{\eta_0,\eta_+,\eta_-}$ subordinate to this division, then we can divide $P(h)$ accordingly, too. More precisely, let $P_i = \frac{1}{2\pi}\int_{\Real}\eta_i e^{-\mathrm{i}sh}e^G\mathrm{d}s$, so that $P = \sum_{i\in \set{0,\pm}}P_i$. For $i\neq 0$, $P_i$ is the Fourier transform of a smooth function, and therefore decays exponentially with $h$. Hence, a power-law decay must originate from $P_0$, if there is one at all \footnote{The function $G(s)$ itself is not smooth, but contains a $|s|^3$ term in its series expansion near the origin, as may be visible from a Mellin transform argument; see appendix \ref{appendix:Mellin transform}.}.
Thus, $P(h) \sim P_0(h)$ as $h\to\infty$. The advantage of adding $\eta_0(s)$ is that it vanishes outside $I_0$, and one thus may expand the exponential as before. After expanding, $\eta_0$ may be removed and the integration limits restored to $\pm \infty$, accruing only exponentially small errors. This argument is similar to that presented in \cite{Jones1966}, in the context of the method of stationary phase.

\subsection{The importance of interferences}

Equation \eqref{eqn:h^-4 power law} implies that the probability for observing a large value of $h$ is dominated by the nearest neighbor. To see this, assume that the individual strains $h_i(t)$ are fully incoherent. As a result, cross-terms vanish and the total \emph{intensity} $L$ is the sum of the individual intensities $h_i^2$,
\begin{equation}
L \equiv \sum_{\textrm{source}~i}\overline{h_i^2},
\end{equation}
as one would expect for incoherent electromagnetic radiation. Here, $\overline{h_i^2}$ is the average of $h_i^2(t)$ over the duration of the experiment.
In analogy with the ionizing background produced by quasars, we have \citep[e.g.,][]{MeiksinWhite2003}
\begin{equation}
  P(L) \overset{L\to\infty}{\sim} L^{-5/2} \;.
\end{equation}
This power-law behavior is known to arise from the nearest neighbor \cite{Vincent2014}. Similarly, bearing in mind that $P(h^2)=P(h)/h$, the large-$h$ asymptotic scaling (\ref{eqn:h^-4 power law}) implies
\begin{equation}
P(h^2) = P(h)/h \overset{h\to\infty}{\sim} h^{-5} \sim (h^2)^{-5/2},
\end{equation}
which demonstrates our assertion.

The distribution $P(L)$ can be computed using the technique outlined above (cf. \cite{MeiksinWhite2003}) upon substituting
\begin{equation}
\label{eq:gtL}
    g(t) = \frac{A^2 r_0^2 e^{-2r/r_0}}{2r^2}
\end{equation}
in the expression of $\tilde p(q;A,\omega;r)$. We included a factor of $\frac{1}{2}$ so that eq.~\eqref{eq:gtL} is precisely the time average of eq.\eqref{eq:gth}.
Consequently, the dependence on the random phase $\varphi$ is trivial, and one is left with
\begin{equation}
    G(s) = 3\int\mathrm{d}b\,\phi(b) \int_0^{R/r_0}\mathrm{d}\tau\,\tau^2\bigg[\exp\!\left(\mathrm{i} s b \frac{e^{-2\tau}}{2\tau^2}\right)-1\bigg]\;,
\end{equation}
where $s\equiv q/L_c$ and $b\equiv A^2/L_c$. We choose the normalization to be $L_c\equiv h_c^2$, such that $\lambda L_c \equiv \langle L\rangle$.
The function $G(s)$ gives $P(L)$ upon a Fourier transformation.

The bottom panel of figure \ref{fig:example of a special case calculation} displays $P(L)$ along with $P(h^2)$ for the comparison. While the two distributions exhibit the same power-law behavior at high-$L$, as explained above, there is a stark difference at low $L$.
This emphasizes the crucial role that interference, both constructive and destructive, play in the determination of the observed time series.
Note that the \emph{means} of both distributions are equal, $\lambda L_c=\lambda h_c^2$, reflecting the fact that the variance of independent random variables is additive. In other words, a measurement of $\langle h^2\rangle$ (that is, $\rho_\text{gw}$ for a realistic GW background) does not provide any information about the (in)coherence of the signal.
As a rule of thumb, the distribution is well approximated by a Gaussian for $|h|\ll \sqrt{\lambda}h_c$ (many sources contribute to the observed strain), and by power-law tails for $|h|\gg \sqrt{\lambda}h_c$ (a single, nearby source dominates the signal).

\section{Stochastic Background from Binaries}
\label{sec:physical case}

Having understood the salient features of the PDF of the observed strain in a simplified case, we now turn our attention to a more realistic source of the SGWB: binary mergers of any combination of white dwarfs, neutron stars or black holes.
All, but the final stages of such systems, may be described by a Keplerian orbit perturbed by the gravitational radiation reaction \cite{Maggiore}.

\subsection{Basic relations}

We approximate the sources as Keplerian throughout, and consider circular orbits -- an assumption justified by the circularizing effect of gravitational-wave emission. As a result, the detector measures a certain linear superposition of the two polarizations $h = F_+h_+ + F_\times h_\times$ with wave-forms given by \cite{Maggiore}
\begin{align}
  & F_+h_+(\mathfrak{t}) = h_0(\mathfrak{t})\frac{1+ \cos^2 i}{2}\cos(2\theta)\cos\phi(\mathfrak{t}) \\ &
  F_\times h_\times(\mathfrak{t}) = h_0(\mathfrak{t})\cos i\sin(2\theta)\sin\phi(\mathfrak{t}),
\end{align}
where $\theta, i$ describe the orientation of the binary relative to the detector, $\mathfrak{t} = t_{\textrm{coal}} - t$ is the time to coalescence as measured by the observer (or at the source, as it is co-moving) and
\begin{align}
  & \phi(\mathfrak{t}) = -2\left(\frac{5G(1+z)M_c}{c^3}\right)^{-5/8}\mathfrak{t}^{5/8} + \varphi, \\ &
  h_0(\mathfrak{t}) = \frac{4}{d_L(z)}\left(\frac{G(1+z)M_c}{c^2}\right)^{5/3}\left(\frac{\pi f^\textrm{obs}_\textrm{gw}(\mathfrak{t})}{c}\right)^{2/3}, \\ &
  f^\textrm{obs}_\textrm{gw}(\mathfrak{t}) = \frac{1}{\pi}\left(\frac{5}{256\mathfrak{t}}\right)^{3/8}\left(\frac{G(1+z)M_c}{c^3}\right)^{-5/8}.\label{eqn:gw frequency}
\end{align}
Here,
\begin{equation}
    M_c = \frac{(m_1m_2)^{3/5}}{(m_1+m_2)^{1/5}}
\end{equation}
is the chirp mass, $\varphi$ is a (random) phase and $z$ is the source redshift.
Furthermore, the time to coalescence reads
\begin{equation}
  \mathfrak{t} = \tau_0(M_c,T) - (t_{0,\ret}(r) - t_*),
\end{equation}
where $t_*$ is time the formation time of the binary,
\begin{equation}
t_{0,\ret}(r) = \int_{0}^{\eta_0-r/c}a(\eta)\mathrm{d}\eta
\end{equation}
is the retarded age of the Universe, and $\tau_0(M_c,T)$ is the lifetime of a binary -- the time it would take until both members collide; assuming gravitational-wave emission solely, we have $\tau_0(M_c,T) = 5c^5T^{8/3}(GM_c)^{-5/3}$ \cite[p. 171]{Maggiore}. We set the present-day scale factor to unity.

Since the phase $\varphi$ is random, the average of $\exp(\mathrm{i}q h(t))$ over $\varphi$ returns a Bessel function for $\tilde p$, as before.
Including an exponential decay with attenuation length $r_0$ (to which $d$ is set), this implies
\begin{equation}\label{eqn:single source Fourier}
  \tilde{p}(q;\xi;\mathbf{r}) = J_0\!\left(\sqrt{F_+^2 + F_\times^2} h_0(\mathfrak{t},z(r)) qe^{-r/r_0}\right) \;.
\end{equation}
The redshift dependence induces a dependence on the co-moving distance $r$ through the Friedmann equations.

Next, $\tilde{p}$ must be further averaged over the sphere $\mathcal{S}^2$ as well as all the model parameters $\xi$ in order to get $\psi$. But first of  let us stress that the integral over $t_\textrm{coal}$ is equivalent to an integral over the `starting time' and, as such, it must be weighted by the density of binary progenitors (which we take to be proportional to the star formation rate).

Finally, the large-$h$ asymptotics of $P(h)$ can be obtained upon separating the argument of the Bessel function into an `amplitude' part, $A(r)$, multiplied by a `propagation' part, $\exp(-r/r_0)/d_L = \exp(-r/r_0)a(r)/r$. Following the procedure outlined in \S \ref{sec:simple case}, one obtains the same asymptotic expansion at large $h$. Namely (see also appendix \ref{appendix:Mellin transform},
\begin{equation}\label{eqn:h^-4 power law physical}
  P(h) \sim \frac{2N_0\langle b(\tau = 0)^3\rangle}{\pi h^4} \;,
\end{equation}
where $b(\tau = 0)$ is the normalized amplitude evaluated at $\tau = 0$. The coefficient $b$ is evaluated at $\tau = 0$, because, in passing through eq. \eqref{eqn:equation18}, $b$ is now a function of $\tau$, which we Taylor-expand about $0$. The terms above zeroth order contribute higher powers of $h$ in the denominator, and are thus neglected.

\subsection{Probability measure for the source parameter space}
\label{subsec: GW parameters}

Before delving into the calculation of the SGWB characteristic function, let us remark on the probability measure we adopt for the model parameters $\xi=(t_*,M_c,T,M_{\textrm{gal}})$. While we attempt to use physically plausible rates, our main goal is to illustrate how our formalism can be used to calculate the PDF of the SGWB produced by binary mergers. Therefore, even though the rates and parameters chosen here were picked so as to simplify this task, we stress that the procedure laid down in this section may be used for any choice of rate functions, such as more accurate estimates based on stellar evolution simulations and past GW merger events \cite{Bocoetal2019,LIGOVirgo2018b}.

To demonstrate our technique, we adopt the so-called `reference model' of \citet{Cusinetal2019}, which we summarize below for the readers' convenience. Let us approximate the probability density that a binary of compact stars forms at cosmic time $t_*$ with semi-major axis $a$ and chirp mass $M_c$ as the product distribution
\begin{align}
  \phi(t_*,M_c,T,M_\textrm{gal}) &= R_*(t_*) \phi(M_\text{gal}|t_*)\phi(M_c|t_*,M_\text{gal})
  \nonumber \\
  & \qquad \times \frac{\ln\left(a_{\max}/a_{\min}\right)}{a} \;.
\end{align}
$R_*(t_*)$ is a probability distribution which accounts for the redshift evolution of the star formation rate, which we model as
\begin{equation}
  R_*(z) = (1+z)e^{-z^2/(2\sigma^2)}\left(\sigma^2+\sqrt{\frac{\pi}{2}} \sigma\right)^{-1}\;.
\end{equation}
Here, $\sigma = \sqrt{6}$, so that $R_*$ peaks at redshift $z = 2$ \cite{Smitetal2012}. $R_*(z)$ can be converted into $R_*(t_*)$ using the redshift-to-cosmic-time relation. The conditional probabilities $\phi(M_\text{gal}|t_*)$ and $\phi(M_c|t_*,M_\text{gal})$ will be discussed shortly.
Finally, the $1/a$ probability density for the initial orbital period is derived from \"{O}pik's law \cite{Oepik1924}, which is a reasonable approximation to the observed Galactic period distribution over a fairly large range of periods \cite{DucheneKraus2013}.
We choose the bounds $a_{\min} = 0.014~\textrm{AU}$ and $a_{\max} = 4000~\textrm{AU}$ as in \cite{Cusinetal2019}. They translate into limits on the initial period $T$ (by Kepler's third law) using the masses of the binary components.

The knowledge of $T$ allows us to compute the binary lifetime $\tau_0$. Then, $G(s)$ is obtained from an integration over $t_*, M_c,T,M_\textrm{gal}$. The non-trivial integration limits are as follows: $T$ lies between $T_{\max}$ and $\max\set{T_{\min},\tau_0^{-1}(t_{0,\textrm{ret}} - t_*)}$, while $t_*$ runs from $0$ to $t_{0,\textrm{ret}}(r)$. The reason for this choice of integration limits is twofold: binaries which merge such that the signal from the merger event reaches the detector (origin) before $t_0$ do not contribute to the gravitational-wave signal at $t_0$; and neither do binaries which form at $(t_*,r)$ such that the signal sent at their birth does not reach the detector in time.

To model the chirp-mass distribution $\phi(M_c|t_*,M_\text{gal})$, suppose that the initial masses $m_1$ and $m_2$ of the two binary members have both broken power-law densities, $\phi(m) = Cm^{-\alpha}$ with $\alpha$ dependent on $m$. We choose a Kroupa mass function \cite{Kroupa2002,Binney} (in the mass-range we consider $\alpha=2.7$), rather than a Salpeter mass function as in \cite{Cusinetal2019}. This is only difference with their model.

The SGWB is mainly produced by remnants of stellar evolution, whose final mass is related to $m_1,m_2$ by the so-called `initial-to-final mass function' $\mu(m,Z)$, which depends on the metalicity $Z$. Following \cite{Cusinetal2019}, we adopt the functional form of the delayed model presented in \citet{Freyeretal2012} (here, all masses are in solar mass units):
\begin{widetext}
\begin{equation}
  \mu(m,Z) = \begin{cases}
               1.3 , & \mbox{if } m \leq 11 \\
               1.1 + 0.2e^{(m-11)/4} - (2 + Z)e^{2(m-26)/5}, & \mbox{if } 11 < m \leq 30 \\
               \min\set{33.35 + (4.75 + 1.25Z)(m-34),m-\sqrt{Z}(1.3m-18.35)}, & \mbox{otherwise}
             \end{cases}
\end{equation}
\end{widetext}
The metalicity depends on the cosmic time at which the system formed, a dependence which we model (following again \cite{Cusinetal2019}) using the fit of \citet{Maetal2016} for the gas density (from which the stars formed):
\begin{equation}\label{eqn:metallicity as function of redshift}
\begin{aligned}
  \log_{10}\left(\frac{Z(z,M_{\textrm{gal}})}{Z_\odot}\right) & = 0.35\left[\log_{10}\left(\frac{M_{\textrm{gal}}}{M_\odot}\right) - 10\right] \\ & + 0.93e^{-0.43z} - 1.05 \;.
\end{aligned}
\end{equation}
The redshift is $z$ then, again, converted to cosmic time $t_*$ using the cosmic-time-to-redshift relation.

All this leads to a chirp mass distribution
\begin{equation}
\phi(M_c|t_*,M_\text{gal})\equiv \phi(M_c|Z(t_*,M_\text{gal})
\end{equation}
given by
\begin{align}
  \phi(M_c|Z) & = \iint \mathrm{d}m_1\mathrm{d}m_2\, \phi(m_1)\phi(m_2)\, \\&\quad
  \times\delta_\text{D}\!\!\left(M_c - \frac{(\mu(m_1,Z)\mu(m_2,Z))^{3/5}}{(\mu(m_1,Z)+\mu(m_2,Z))^{1/5}}\right) \nonumber \;,
\end{align}
where $\delta_\text{D}(x)$ is the Dirac delta-function.

The last ingredient is $\phi(M_\textrm{gal}|t_*)$, which we model using the halo mass function of \citet{Tinkeretal2016} and under the assumption that the total stellar mass in a galaxy is proportional to its halo mass.

The knowledge of the functions $R_*(t_*)$, $\phi(M_\textrm{gal},t_*)$, $\phi(M_c|t_*,M_\text{gal})$ and $\phi(a) \propto 1/a$ fully specifies the PDF of $\phi(\xi)$, up to an overall rate encapsulated in the value of $N_0$ which we are free to specify.

\subsection{Determination of $N_0$}

$N_0$ is the average number of overlapping sources in the time domain data, whose frequencies are in the detector's observing band. While $N_0$ could be as large as, say, $10^6- 10^9$ if we were observing the stochastic background across all frequencies, it is of order $\mathcal{O}(10)$ only if frequencies are restricted to $f_\text{gw}^\text{obs}\gtrsim 1$ Hz \citep{LIGOVIRGO2018a}.
Here and henceforth, we will match our predictions to the LIGO's frequency band. This restricts the source parameters via equation \eqref{eqn:gw frequency}, given by $f_{\min} \leq f^\textrm{obs}_\textrm{gw} \leq f_{\max}$, where $f_{\min} = 5 ~\textrm{Hz}$ and $f_{\max} = 5000~\textrm{Hz}$. Imposing such a constraint on $f_\textrm{gw}^\textrm{obs}$ through an additional Heaviside function in the definition of $G(s)$ (see below) affects the determination of $N_0$.

To determine $N_0$ (and, thereby, the absolute normalisation of the star formation rate $R_*(z)$ in our model), we use the approximate relation
\begin{equation}
N_0 \sim R_{\textrm{merger}}\cdot V\cdot \bigl\langle \mathfrak{t}(f_{\min})\bigr\rangle \;,
\end{equation}
where $V$ is the comoving volume of the observable Universe (because we assume the detector to be noiseless in the observing frequency band), $R_{\textrm{merger}}$ is the present-day merger rate (inferred from the data), and $\bigl\langle \mathfrak{t}(f_{\min})\bigr\rangle$ is the average amount of time spent by a binary in the detector's observing band (which, in addition to $f_\text{min}$ and $f_\text{max}$, sensitively depends on the physics of GWs).

For $R_\text{merger}\sim 2000\, {\rm Gpc}^{-3}{\rm yr}^{-1}$, $V\sim 10\, {\rm Gpc}^3$
and $\bigl\langle\mathfrak{t}(f_{\min})\bigr\rangle\sim 200\,{\rm s}$, we obtain $N_0\sim 10$, in agreement with the value of $N_0=15$ calculated by ref. \cite{LIGOVIRGO2018a} in the LIGO frequency band \footnote{The rates in ref. \cite{LIGOVIRGO2018a} are slightly lower than those in the more recent refs. \cite{LIGOVirgo2018b,LIGOvirgo2019PRX}, but consistent with them, and the value of $N_0$ changes only by an order-unity constant, which does not alter any of our conclusions.}. We shall adopt $N_0=15$ as our fiducial value.

\subsection{Distribution of observed strain fluctuations}
\label{subsec:SGWBdist}

All that remains is to evaluate $G(s)$, i.e. to compute
\begin{widetext}
\begin{equation}\label{eqn:G}
\begin{aligned}
  G(s) & = 3\int_{r_{\min}/r_0}^{R/r_0} \mathrm{d}\tau\, \tau^2\int_{\mathcal{S}^2} \mathrm{d}\Omega \int \mathrm{d}M_c \int \mathrm{d}T \int \mathrm{d}t_* \int \mathrm{d}M_\textrm{gal}\, \varphi(t_*,M_c,T,M_\textrm{gal}) \Big(\tilde{p}(s;\xi;\tau,\theta,\phi) - 1\Big) \\ \quad  &
  \times \Theta\left[(f_\textrm{gw}^\textrm{obs} - f_{\min})(f_{\max} - f_\textrm{gw}^\textrm{obs})\right],
\end{aligned}
\end{equation}
\end{widetext}
where $\Theta$ is Heaviside's function.
We perform this integral numerically assuming $R=14$ Gpc and $r_0=10$ Gpc (see appendix \ref{appendix:integral} for details).
For our fiducial cosmology, 10 and 14 Gpc correspond to the comoving radial distance to redshift $z\simeq 12$ and to the last scattering surface, respectively.
The result is plotted in figure \ref{fig: G physical r0=0.01}.

\begin{figure}
  \centering
  \includegraphics[width=0.45\textwidth]{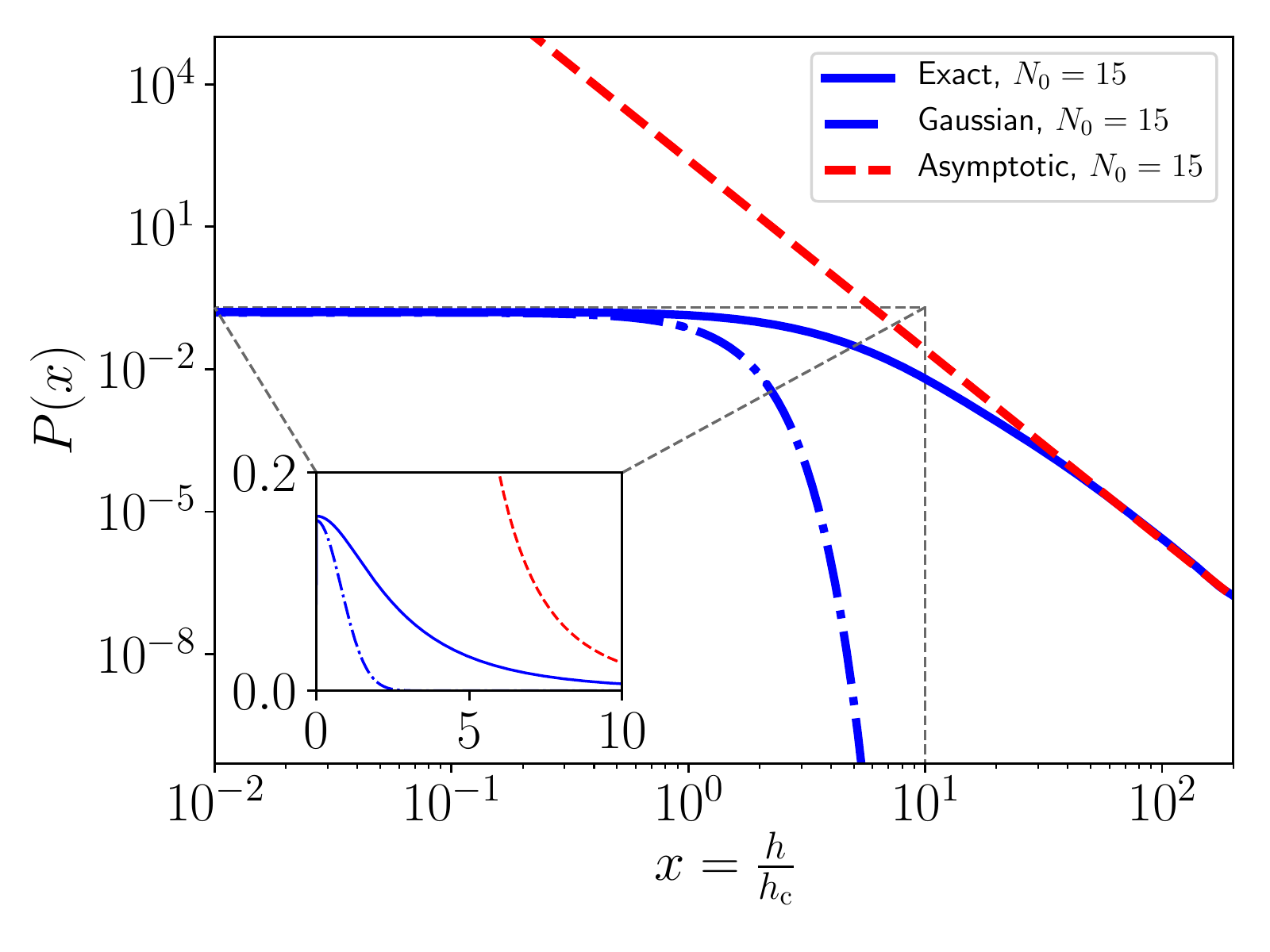}
  \caption{The probability density function of the cumulative SGWB strain from binaries for a mean number of sources $N_0 = 15$ (solid blue) within one attenuation volume (see text for details). There is excellent agreement with the large-$h$ asymptotic prediction in equation \eqref{eqn:h^-4 power law physical} (red dashed). The Gaussian approximation of the probability density function, given by the second order expansion of $G(s)$, is also shown for comparison (blue dotted dashed).}
  \label{fig: P(h) physical}
\end{figure}

\begin{figure}
\begin{center}
    \includegraphics[width = 0.45\textwidth]{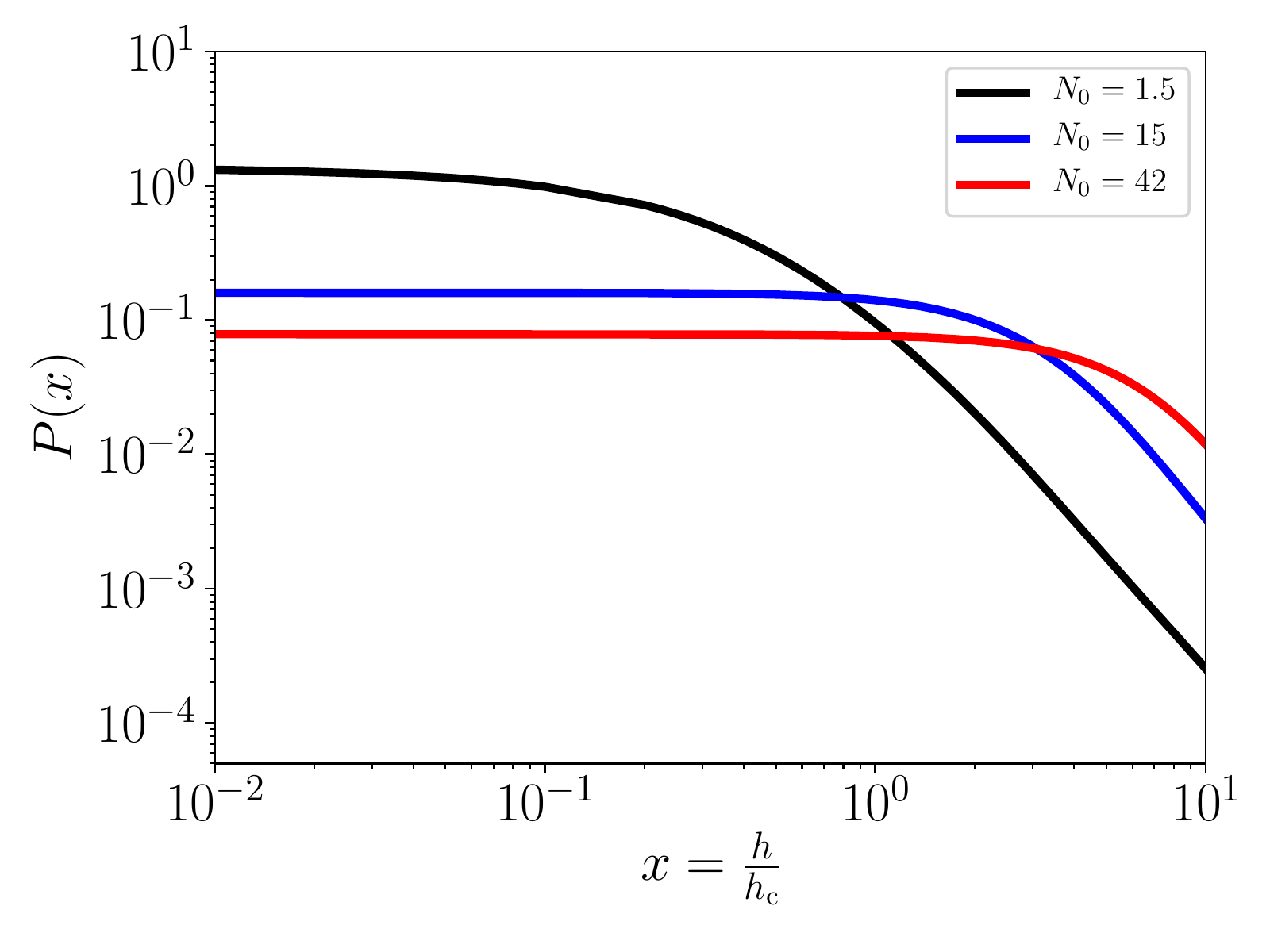}
    \caption{The probability density function $P(h)$ of the strain $h$ for different values of $N_0$ The black line corresponds to our fiducial choice of $N_0 = 15$, while red and blue are $N_0 = 1.5$ and $N_0 = 42$ respectively. Note that $N_0$ always refers to the attenuation volume, which we give in appendix \ref{appendix:integral}.}
    \label{fig:pdf_shaded}
    \end{center}
\end{figure}

The distribution $P(h)$, computed from an inverse Fourier transform of $\exp(N_0G)$, is shown in figure \ref{fig: P(h) physical} assuming $N_0 = 15$.
We find a strain normalization of $h_c = 2.06\times 10^{-25}$. Here again, there is an excellent agreement between the exact numerical result (solid curve) and the asymptotic expression (dashed line) in the large-$h$ limit. We have also overlaid a Gaussian distribution (dotted-dashed curve) with variance $\langle h^2\rangle$ equal to that of the full distribution. This emphasizes that a Gaussian is a bad approximation over the range of strain values considered here owing to the low number of sources.
As $N_0$ increases, the transition to the $h^{-4}$ power-law tail moves to larger values of $h$, so that the Gaussian approximation improves (at fixed value of $h$). In figure \ref{fig:pdf_shaded}, we show the effect of varying $N_0$ on the probability $P(h)$ of measuring a squared strain $h$.

To conclude this section, recall that the variance of observed strain fluctuations is related to the density parameter $\Omega_\textrm{gw}(f)$ through
\begin{equation}
\langle h^2(t)\rangle = \frac{8G}{\pi c^2} \rho_\text{crit}\int_0^\infty\!\mathrm{d}\ln f\,f^{-2}\Omega_\text{gw}(f) \;.
\end{equation}
The unequal time correlator $\langle h(t) h(t')\rangle$ encodes additional information on, e.g., whether the SGWB produced by binary mergers of compact stars falls in the ``continuous'', ``shot noise'' or ``pop-corn'' regime \cite{Regimbau:2011rp}. This depends on the ratio between the duration of events and time interval between successive events. At the distribution level, quantifying the correlation structure of time series data would amount to calculating the 2-point PDF $P\big(h(t),h(t')\big)$ (and higher order statistics). This ought to be relatively straightforward, albeit beyond the scope of this paper.

\section{The confusion background}
\label{sec:confusion background}

So far, we have computed the distribution $P(h)$ of time domain strain fluctuations measured in the frequency domain $f_{\min} \leq f^\textrm{obs}_\textrm{gw} \leq f_{\max}$ by an idealized, noiseless detector without any signal post-processing (i.e. source identification etc). In this Section, we will model the outcome of a realistic analysis in which bright sources significantly above the detector sensitivity are subtracted from the time-domain series. The resulting distribution, which we denote $P_r(h)$, will characterize the so-called "confusion background" of unresolved sources \citep{Timpanoetal2006,Regimbau:2009rk}.

\subsection{Bright source subtraction}

When a binary system has a large apparent brightness -- as is the case of nearby sources -- its signal can rise well above the residual noise rms variance $\sqrt{\langle n^2\rangle}$. As a result, it can be resolved as an individual event and subtracted from the SGWB time series if it does not significantly interfere with other signals. In this case, it ceases to contribute to the SGWB as defined above.
In practice, the identification, modeling and subsequent removal of bright sources are quite complicated. Therefore, we shall consider here the following simplified implementation.

A black-hole binary produces waves whith frequency rising in time, until its components coalesce. The detector observes the merging while the frequency $f^\textrm{obs}_\textrm{gw}$ of the gravitational waves lies between $f_{\min}$ and $f_{\max}$.
The signal-to-noise ratio of the detector is \cite{Maggiore}
\begin{equation}
\begin{aligned}
    \left(\frac{S}{N}\right)^2 & = 4\int_{f_{\min}}^{f_{\max}} \mathrm{d}f \frac{\abs{\tilde{h}(f)}^2}{S_n(f)} \\ &
    = \frac{1}{3\pi^{4/3}}\frac{c^2}{d_L(z)^2}\left(\frac{GM_c(1+z)}{c^3}\right)^{5/3}\\&
    \times \int_{f_{\min}}^{\min\set{2f_\text{ISCO},f_{\max}}}\mathrm{d}f \frac{f^{-1/3}}{f^2S_n(f)}
\end{aligned}
\end{equation}
where $f_\text{ISCO} = 2200~\textrm{Hz }\times \frac{M_\odot}{m_1+m_2} $, and $S_n(f)$ is the detector's noise spectral density \footnote{\url{https://dcc.ligo.org/LIGO-T1800044/public} \cite{Barsottietal2018}}. For convenience, let us introduce
\begin{equation}
    I = \int_{f_{\min}}^{440 ~\textrm{Hz}}\mathrm{d}f \frac{f^{-1/3}}{f^2S_n(f)} \;.
\end{equation}
Because of the $f^{-7/3}$ power in the integrand, changing the upper limit of $I$ from the maximum possible value of $2f_\text{ISCO}$ in the model used here, $\sim 440$ Hz, to its minimum, has a small effect on its value. Thus, requiring the source SNR to be more than a certain value $n \sigma$ amounts to requiring that
\begin{equation}\label{eqn: bright condition}
    \frac{(M_c (z+1))^{5/3}}{d_L^2} \geq \frac{3\pi^{4/3}c^3k^2}{G^{5/3}I} \approx 3.3n^2\times 10^{-7} M_\odot^{5/3} ~\textrm{Mpc}^{-2}\;.
\end{equation}
If this inequality is satisfied, the source is deemed \emph{bright}, and its signal is removed from the data provided that its time to coalescence $\mathfrak{t}$ is smaller than $5$ years (so that it merges during a 5 years observational run).
In practice, the bound on the SNR may be formally expressed as a bound on a function of the source parameters.
The latter is then inserted as a Heaviside function into the integrand of equation \eqref{eqn:G}, thereby ensuring that the condition \eqref{eqn: bright condition} is not satisfied by the sources making up the confusion background.

\subsection{Distribution}

\begin{figure}
  \centering
  \includegraphics[width=0.45\textwidth]{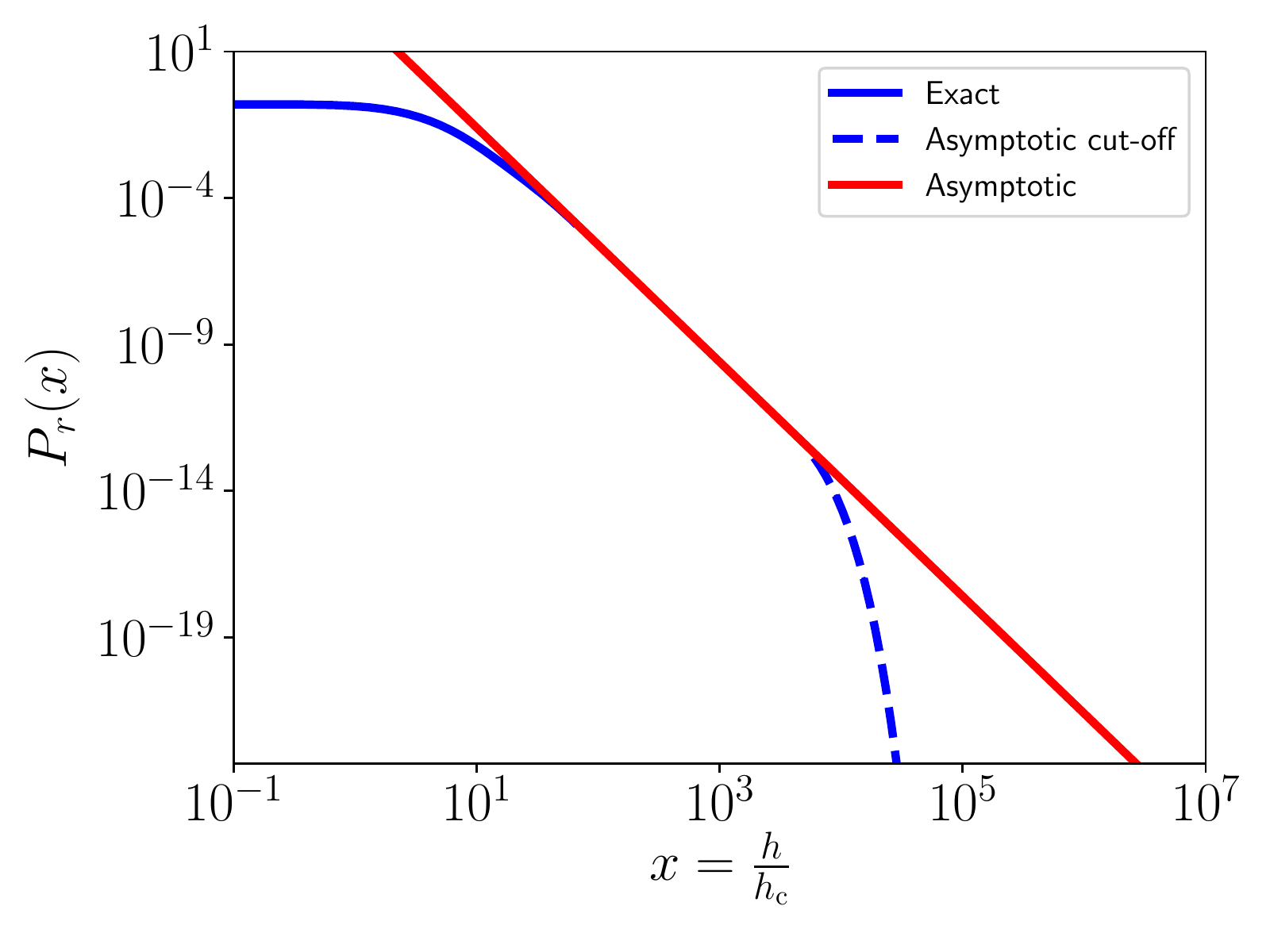}
  \caption{Full shape of $P_r(x)$ with realistic $h_{\textrm{cutoff}}$: power-law and exponential for $x\ (h)\to\infty$. Details of the computation can be found in appendix \ref{appendix:exponential asymptotic}.}
  \label{fig:asymptotics}
\end{figure}

The condition Eq.(\ref{eqn: bright condition}) introduces another strain scale, $h_\textrm{cutoff}$, in addition to $h_c$. Above this scale, typically much larger than $h_c$, one expects an exponential fall-off of $P_r(h)$ (cf. appendix \ref{appendix:exponential asymptotic}).
For an experiment with LIGO characteristics, we obtain indeed $h_\textrm{cutoff} \sim 10^4 h_c\gg h_c$. In this case, we expect three different regimes as shown in Fig.~\ref{fig:asymptotics}: $h \ll N_0 h_c$, where $P_r(h)$ is roughly constant, $h_\textrm{cutoff} \gg h \gg N_0h_c$, where $P_r(h)$ behaves like a power-law, and $h > h_\textrm{cutoff}$, above which $P_r(h)$ decreases exponentially. This emphasizes that the distribution $P_r(h)$ of the confusion background $P(h)$ is very close to $P(h)$ when $h_\text{cutoff}\gg h_c$.
Overall, any cutoff $h_\textrm{cutoff}$ will regularize the moments of $P(h)$ (which are all infinite starting from the third), and give rise to an exponential decline. The variance $\langle h^2\rangle$ remains weakly sensitive to the value of $h_\textrm{cutoff}$ (i.e. it varies at most by $\sim$10\%) so long as $h_\textrm{cutoff}\gg h_c$.

\begin{figure}
\begin{center}
    \includegraphics[width = 0.45\textwidth]{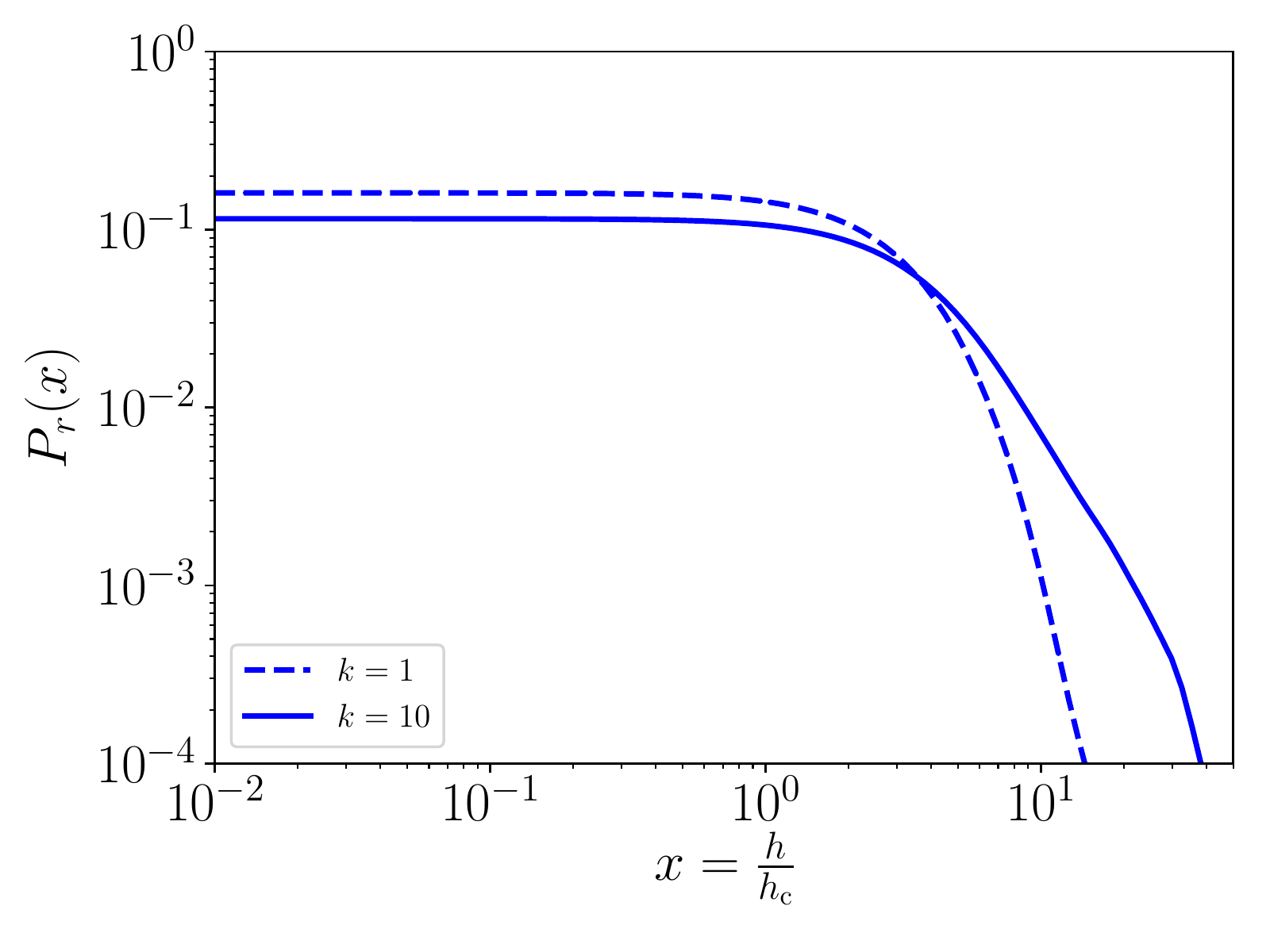}
    \caption{The distribution of the confusion background $P_r(h)$ for two different cutoff SNR$>k h_c$, with $k=1$ and 10, for the bright source subtraction.}
    \label{fig:pdf_cut}
    \end{center}
\end{figure}

To illustrate the impact of bright source subtraction when $h_\text{cutoff}$ is comparable to $h_c$, we consider an hypothetical detector with sensitivity comparable to $h_c$ (and, thus, much better than LIGO). Upon removing sources which are $k$ times brighter than $h_c$, i.e. SNR$> k h_c$, with $k=1$ or 10, we obtain the distribution $P_r(h)$ of the confusion background displayed in \cref{fig:pdf_cut} (the evaluation of the Fourier transform becomes computationally challenging for large values of $x$). One can spot a cutoff around $x\sim k$ at which the slope of the distribution changes. This is particularly clear for the case $k=10$ because the intermediate, $h^{-4}$ power-law regime is still visible. For the case $k=1$, this intermediate regime is not apparent because $P_r(h)$ quickly transitions to the steeper fall off, rendering the full shape of $P_r(h)$  Gaussian-like. This agrees with the numerical findings of \cite{Timpanoetal2006}, who found that $P_r(h)$ becomes closer to a Gaussian when bright sources are removed.
Clearly however, $P_r(h)$ depends on the details of the source subtraction procedure.
Note also that, by definition, a given detector will be sensitive to values of $h$ not significantly below the cutoff $h_\textrm{cutoff}$.
However, hypothetically speaking, one could decide to deem a source as bright only if it is detected at $5\sigma$, but settle for finding the SGWB at $3\sigma$, in which case one could still measure the flat and/or power-law behaviour of $P_r(h)$.

To conclude this discussion, we emphasize that cross-correlations between different detectors can help reducing the experimental noise \cite{Maggiore}. In this case however, it is the probability density $P_r(h^2)$ which is the quantity being measured \footnote{The overlap reduction function is used to take into account the fact that the detectors are not spatially coincident \citep{Flanagan:1993ix,Allen:1997ad}. However, the interpretation of the cross-correlation signal may be less straightforward. We thank the anonymous referee and Amos Ori for pointing this out to us.}. This can help probing the distribution $P_r(h)$ of the confusion background for values of $h$ much smaller than the sensitivity of a single detector.

\section{Discussion and General Properties}
\label{sec:discussion}

The salient features of $P(h)$ we outlined in this paper remain valid in a more general setting -- when $g = A(t)\exp[\mathrm{i}(\omega(t)t +\varphi)]$ --  provided that $A$ and $\omega$ are slowly-varying functions of $t$. As in Sec.~\S\ref{sec:simple case} and \S\ref{sec:physical case}, averaging over the phase $\varphi$ gives a Bessel function:
\begin{equation}
  \tilde{p} = J_0\!\left(qA\frac{de^{-r/r_0}}{d_L(r)}\right) \;.
\end{equation}
For $h_\textrm{cutoff} = \infty$, the dominant contribution arises again from the $\abs{s}^3$ term in $G(s)$, and generates an $h^{-4}$ power-law.
The $-4$ exponent actually reflects the specific $r$-dependence of the GW luminosity distance in General Relativity (GR). Obviously, this exponent may change if the flux does not satisfy the familiar $1/r^2$ law (as is the case in some extensions of GR \cite{Cliftonetal2012,Ishak2018}).
For $h_\textrm{cutoff} < \infty$, the methods of Sec.~\S\ref{sec:confusion background} and appendix \ref{appendix:exponential asymptotic} apply, \emph{mutatis mutandis}.

\subsection{Isotropy}

One critical assumption made throughout this paper is that of the isotropy of the source spatial distribution (the mean number density is allowed to vary along the radial direction), so that our $\langle h^2\rangle$ actually corresponds to the shot noise term discussed in \cite{Jenkinsetal2019b}, rather than the angular power spectra $C_\ell$'s calculated in \cite{Cusinetal2017,Jenkinsetal2018}
\citep[see also][for similar  calculations in the context of cosmological backgrounds]{Geller:2018mwu,Bartolo:2019oiq,Bartolo:2019zvb}.

While isotropy is a reasonable assumption for the main extra-Galactic background sources of LIGO and Virgo \cite{Jenkinsetal2019b}, there is a significant contribution from Galactic white-dwarf binaries to the background that should be observed by {\small LISA} \cite{Maggiore}. The formalism considered here can be extended to include generic clustering \citep[along the lines of, e.g.,][]{Vincent2014} and projection effects induced by inhomogeneities (peculiar velocities, gravitational redshift etc.) into the calculation of $G(s)$ \cite[see for instance][]{Bertacca:2019fnt}.

In the specific case of contributions from our Galaxy, eq. \eqref{eq:pdfindividualsource} will have to be amended to include a number density which depends on the sky direction in accordance with the Galactic density profile (approximately a disk, see \citep{Conneelyetal2019}). The SGWB will then be the superposition of an extra-Galactic part $h_{\textrm{ext}}$ (studied in this paper), and a Galactic part $h_\textrm{Gal}$. Since both are independent, the characteristic function of $h = h_{\textrm{extra}} + h_\textrm{Gal}$ is the product of their characteristic functions. One can still denote the characteristic function of $h_\textrm{Gal}$ by $\exp(N_\textrm{Gal}G_\textrm{Gal}(s))$, with $N_\textrm{Gal}$ measuring the mean number of galactic sources across the sky (even if this distribution is neither isotropic nor Poissonian). Therefore, the function $G(s)$ should also separate into two parts: $N_{\textrm{tot}}G(s) = N_0G_\textrm{ext}(s) + N_\textrm{Gal}G_\textrm{Gal}(s)$. If $N_\textrm{Gal} \gg 1$, Laplace's method (this time treating $N_\textrm{Gal}$ as the big parameter, not $h$) ensures that, for all values of $h$ but the very largest, where the approximations of appendix \ref{appendix:exponential asymptotic} applies, the dominant contribution to $G_\textrm{Gal}$ is its $O(s^2)$ term in its Maclaurin series. Thus, one expects that for large $N_\textrm{Gal}$, $h$ is well-approximated by a sum of $h_\textrm{ext}$, whose distribution we have described here, and a Gaussian random variable, provided that $h$ is not too large.

\subsection{Frequency domain}
\label{sec:frequencydomain}

All of the above is true for the entire gravitational-wave amplitude integrated, so to speak, over all frequencies. However, we can resolve the probability distribution for each frequency.
A signal processing in the frequency domain is also desirable (especially when $\langle n^2\rangle$ is of order $\lambda h_c^2$ or larger) because the detector noise usually has a very specific frequency dependence.

Let $\tilde{h}(f) = \int\!\mathrm{d}t\, e^{2\pi \mathrm{i} tf}h(t)$ be the Fourier amplitudes. For $N$ sources, we have
\begin{equation}
  \tilde{h}(f) = \sum_{k=1}^{N}\tilde{h}_k(f)\;.
\end{equation}
The individual Fourier modes \footnote{One could also consider a wavelet decomposition.}
$\tilde h_k(f)$ also are independent by the assumptions stated in \S \ref{sec:general}, with the caveat that they are complex random variables. Treating $\tilde h(f)$ as a complex variable, the single source characteristic function $\psi(q)$ should be defined as
\begin{equation}
\psi(q) = \left\langle \exp\!\Big[\mathrm{i}\, \Re\big(q\tilde g(f)\big)\Big]\right\rangle\;,
\end{equation}
where $q$ is now complex, $\tilde g(f)$ denotes the Fourier transform of $g(t)$ and $\Re(z)$ is the real part. Alternatively, since $\tilde h(f)$ must satisfy the reality condition (because $h(t)$ is real), we can restrict ourselves to statistics of the real part $\Re\big(\tilde h(f)\big)$ of the Fourier amplitudes without loosing any information. In this case, the methods of Sec.~\S\ref{sec:general} applies exactly provided that $g(t)$ is replaced by
\begin{equation}
  \tilde{g}(f) = \frac{1}{T}\int_0^T\! \mathrm{d}t\, \cos(2\pi ft)\, g(t) \;.
\end{equation}
($T$ expresses the finite duration $T$ of the experiment; frequencies $f<1/T$ are poorly sampled by the data.)

As an illustration, consider the simplified model of Sec.~\S\ref{sec:simple case}. For a general $T<\infty$, the Fourier transform $g$ yields
\begin{align}
    \tilde g(f) &= \frac{A}{2T} \frac{r_0 e^{-r/r_0}}{r} \bigg[\left(\frac{\cos\omega_+}{\omega_+}+\frac{\cos\omega_-}{\omega_-}\right)\sin\varphi \nonumber \\
    &\qquad +\left(\frac{\sin\omega_+}{\omega_+}+\frac{\sin\omega_-}{\omega_-}\right)\cos\varphi\bigg]\;,
\end{align}
where $\omega_\pm = \omega\pm 2\pi f$. Integrating over the random phase $\varphi$, $\tilde p(q,f;\xi;{\bf r})$ becomes, after some manipulations,
\begin{equation}
\label{eq:tildepsimplefreq}
    \tilde p(q,f;\xi,{\bf r})=J_0\!\left(q\frac{Ar_0 e^{-r/r_0}}{r}\alpha(\omega,f,T)\right),
\end{equation}
provided that $\alpha(\omega,f,T)$ is defined as
\begin{equation}
    \alpha(\omega,f,T)=\frac{\sqrt{\omega^2+4\pi^2f^2+\big(\omega^2-4\pi^2f^2\big)\cos(4\pi fT)}}{\sqrt{2\,}\big\lvert\omega^2-4\pi^2 f^2\big\lvert\, T}\;.
\end{equation}
Eq. \eqref{eq:tildepsimplefreq} is identical to eq. \eqref{eq:tildepsimple} except for a frequency-dependent factor of $\alpha(\omega,f,T)$. Our previous arguments remain valid, so the PDF $P(\Re\big(\tilde h(f)\big)=h)$ will be dominated by the nearest neighbor at large $h$. Its asymptotic form thus exhibits the power-law behavior $\sim h^{-4}$, although its overall amplitude is now modulated in accordance with $\alpha(\omega,f,T)$.

\subsection{Central limit theorem}

The reader might wonder whether central limit theorems (CLTs) hold here, especially the {\it classical} CLT that guarantees the {\it pointwise} convergence of a sum of identically distributed variables with {\it finite} variance. In our case however, the variance of a single-source is infinite.

To see this, consider the simplified model of \S \ref{sec:simple case}, where the emitted amplitude $b$ is constant, and let us evaluate $\mathbb{E}\langle h^2\rangle$ ($\mathbb{E}h = 0$), where $\langle \cdot \rangle$ is a phase average. The tail formula for the expectation gives
\begin{align}
  \mathbb{E}\langle h^2\rangle & = \int_{0}^{\infty}\mathrm{d}t ~P(\langle h^2\rangle > t)
  = \int_{0}^{\infty}\mathrm{d}t ~P\!\left(\frac{d^2b^2}{2r^2} > t\right) \nonumber \\
  & = \int_{0}^{\infty}\mathrm{d}t ~P\!\left(r < \frac{db}{\sqrt{2t}}\right)
  = \int_{0}^{\infty}\mathrm{d}t \int_{0}^{\frac{db}{\sqrt{2t}}} \mathrm{d}r \frac{3r^2}{R^3} \nonumber \\
  & = \int_{0}^{\infty}\mathrm{d}t \frac{d^3b^3}{R^3(2t)^{3/2}} \;.
\end{align}
This diverges, of course, at $t = 0$. Therefore, the SGWB does not satisfy the {\it classical} CLT. However, we argue that $P(h)$ still converges (non-uniformly) towards a Gaussian despite the infinite variance of the single source distribution, in agreement with the existence of CLTs for random variables with infinite variance \citep[e.g.,][]{Berry:1941,Esseen:1942,BORGERS2018679}.

To demonstrate this point, let us scrutinize Eq.~\eqref{eqn:approximation for P in terms of Q}. If the series in the argument of the exponential would involve equal powers of $N_0$, $\langle b^2\rangle$ and $s$, i.e. $- N_0 \langle b^2\rangle s^2 + (N_0\langle b^2\rangle)^{3/2}|s|^3+ ...$ (we omit $\mathcal{O}(1)$ series coefficients for clarity), then any change in the value of $N_0$ or $\langle b^2\rangle$ could be absorbed into a redefinition of $s$. Therefore, the shape of $P(h)$ for large $N_0\gg 1$ would be approximately the same as that for $N_0\sim 1$, that is, highly non-Gaussian. The linearity of the series expansion in $N_0$ implies that $P(h)$ eventually converges to a Gaussian in the limit $N_0\to\infty$. However, the rate of convergence strongly depends on the value of $h$ and the statistical properties of the source distribution. In the random walk analogy, the slow convergence manifests itself as anomalous diffusion on a longer timescale than expected from the classical CLT.

Comparing the $s^2$ and $|s|^3$ terms in Eq.~\eqref{eqn:approximation for P in terms of Q} shows that the closer $P(h)$ is to a Gaussian, the smaller is the ratio
\begin{equation}
\frac{\langle b^3\rangle}{\sqrt{N_0}\langle b^2\rangle^{3/2}}  \equiv \frac{s_3(b)}{\sqrt{N_0}} \;,
\end{equation}
where $s_3(b)$ is the skewness of the $b$-distribution only if $\langle b\rangle =0$.
In other words, source distributions with a large $s_3$ lead to a slower convergence rate, in agreement with the Berry-Esseen theorem \citep{Berry:1941,Esseen:1942}.
For illustration, let us evaluate $s_3$ for the initial, power-law mass distribution $\phi(m)$. Assuming $m_\text{min}<m<m_\text{max}$ and $2<\alpha<3$, we find
\begin{equation}
    s_3(m) \approx m_\text{max}^{(\alpha-1)/2}
    \quad\mbox{for}\quad
    m_\text{max}\gg m_\text{min} \;,
\end{equation}
i.e. $s_3(m)$ grows nearly linearly with $m_\text{max}$ for $\alpha=2.7$. As a result, taking a value of $m_\text{max}$ ten times larger implies that $N_0$ should be a hundred times bigger to achieve a similar level of convergence.

Overall, the non-uniformity of the convergence rate is reflected by Eq.~\eqref{eqn:h^-4 power law physical} which we have shown to be valid for any value of $N_0$.
Indeed, we find that for $h \lesssim N_0 h_c$, the Gaussian approximation holds, while for larger values of $h$, the power-law asymptotic is recovered. When there is not bright-source removal, for any \emph{finite} $N_0$, $P(h)$ therefore assumes the shape of a power-law at sufficiently large values of $h$. As $N_0$ increases, the power-law regime is pushed further and further to higher values of $h$.


\section{Conclusion}
\label{sec:conclusions}

In this paper, we have exemplified a formalism to calculate the probability density function (PDF) of a stochastic gravitational-wave background produced by Poisson-clustered compact binaries. A similar approach was considered in the context of the high-redshift ionizing UV background but, to the extent of our knowledge, it is the first time that it is applied to stochastic GW backgrounds, where interference is taken into account in a consistent manner.

In contrast with earlier works on the topic, we provided expressions for the full distribution function $P(h)$, not only its variance (or power spectrum).
We also demonstrated how our approach can be extended to calculate the distribution $P_r(h)$ of the confusion background obtained after the subtraction of bright sources.
Our formalism thus has the advantage of being able to model large deviations, as well as typical fluctuations.
We evaluated the PDF of the stochastic gravitational strain numerically, and derived accurate asymptotic expressions for the large strain tail. The latter turns out to be dominated by the nearest active source. The resulting $h^{-4}$ scaling is, in fact, a universal phenomenon, in-so-far-as it is independent of the particular properties and characteristics of the sources -- so long as they are deterministic, point sources -- and arises solely from the nature of gravity, the four-dimensional nature of space-time and the inverse-square law decay of the wave amplitude. Furthermore, we demonstrated that both $P(h)$ and $P_r(h)$ are generically non-Gaussian as they depends on the {\it moments} (rather than the cumulants) of the source parameter distribution owing to the projection along the line of sight to the observer. As a corollary, one must be careful with the application of central limit theorems because the source variance is infinite. We expect these conclusions to hold also for the distributions of Fourier amplitudes.

Section \ref{sec:physical case} described a calculation of the stochastic background produced by binary mergers of compact stars, assuming Newtonian orbits throughout the binary's evolution in conjunction with other simplifying assumptions on the rates, mass functions and spatial clustering of the sources.
These, however, do  reflect a limitation of the formalism, but rather the authors' wish to simplify the numerics. In fact, $G(s)$ may be calculated much more precisely in an analogous way, using rates and mass-function derived from simulations directly and taking into account inhomogeneities in the spatial distribution of the sources. Finally, our results can be extended in the Fourier domain and, therefore, could be useful for the frequency reconstruction of stochastic GW backgrounds (see, e.g., \citep{Caprini:2019pxz}).
We defer a thorough study of all these issues to future work.

\acknowledgments{We would like to thank Adi Nusser for helpful discussions, and the anonymous referee for a helpful criticism of our paper. Y.B.G. acknowledges support from the Technion Jacobs scholarship. V.D. and R.R. acknowledge funding from the Israel Science Foundation (grant no. 1395/16 and 255/18). V.D. would also like to thank the Munich Institute for Astro- and Particle Physics (MIAPP) of the DFG cluster of excellence ``Origin and Structure of the Universe'', and the Institut d'Astrophysique de Paris (IAP), for their hospitality during the completion of this work. H.B.P. acknowledges support from the Merle Kingsley fund at Caltech.}

\bibliography{distribution}

\onecolumngrid

\appendix

\section{Asymptotic expansion for any $N_0$}
\label{appendix:Mellin transform}

We demonstrate that the power-law asymptotic behaviour in equation \eqref{eqn:h^-4 power law} is true even when $N_0$ is not small. As we are interested in the large $\abs{h}$ limit, the dominant contribution comes from small values of $s$ in $G(s)$, so it is important to expand equation \eqref{eqn:G simple case} around $s=0$. We do this using a Mellin transform method (see appendix A.2 of \cite{BarryHughes1995}): let
\begin{equation}
  \bar{G}(\mu) = \int_{0}^{\infty}\mathrm{d}s\,G(s)s^{\mu -1}
\end{equation}
be the Mellin transform of $G$. The integral converges for $0 > \Re (\mu) > -2$, and may be performed analytically \cite{DLMF} to give
\begin{equation}
  \bar{G}(\mu) = 3\int \mathrm{d}b\,\phi(b)\int_{0}^{R/r_0}\mathrm{d}\tau\, \tau^2 \left(\frac{be^{-\tau}}{\tau}\right)^{-\mu}\frac{\textrm{cosec}\!\left(\frac{\mu \pi}{2}\right)}{2^{1-\mu}\Gamma^2\left(1-\frac{\mu}{2}\right)}\;.
\end{equation}
Integrating further over $\tau$, we arrive at
\begin{equation}\label{eqn:Mellin transform of G}
  \bar{G}(\mu) = 3\int \mathrm{d}b\,\phi(b) b^{-\mu}(-\mu)^{-(3+\mu)}\frac{\textrm{cosec}\!\left(\frac{\mu \pi}{2}\right)\Gamma(3+\mu)}{2^{1-\mu}\Gamma^2\left(1-\frac{\mu}{2}\right)}\;.
\end{equation}
This formula can be analytically continued to any complex $\mu$, except for $\mu = 0,\pm2, \pm4, \pm6,\ldots$ (the cosecant's poles) and $\mu = -3,-4,-5,\ldots$ (the numerator gamma function's poles). The Mellin transform inversion theorem implies that, for $c = \Re(\mu)$ between $0$ and $-2$, we can write
\begin{equation}
  G(s) = \frac{1}{2\pi \mathrm{i}}\int_{c-\mathrm{i}\infty}^{c+i\infty}\mathrm{d}\mu\, s^{-\mu}\bar{G}(\mu)\;.
\end{equation}
Now, using the analytical continuation of equation \eqref{eqn:Mellin transform of G} to shift the integration contour all the way to the left. Therefore, all the negative integer poles of equation \eqref{eqn:Mellin transform of G}, starting at $-2$, contribute to the small $s$ expansion of $G(s)$. The first three terms come from $\mu = -2,-3,-4$, \emph{viz.}
\begin{equation}\label{eqn:G Mellin small positive s}
  G(s) \sim -\frac{3\langle b^2\rangle}{8\pi}s^2 + \frac{\langle b^3\rangle}{3\pi}s^3 + \frac{24 \gamma_E -27+12 \ln2}{64 \pi }\langle b^4\rangle s^4 + \ldots \;,
\end{equation}
where $\gamma_E$ is Euler's constant.

The Mellin transform is only affected by positive values of $s$, so the above equation is correct only for small, positive $s$. For negative values, recall that $G(s)$ is an even function of $s$, and extend equation \eqref{eqn:G Mellin small positive s} to negative values evenly, by adding an absolute value to all odd powers of $s$:
\begin{equation}\label{eqn:G Mellin small s}
  G(s) \sim -\frac{3\langle b^2\rangle}{8\pi}s^2 + \frac{\langle b^3\rangle}{3\pi}\abs{s}^3 + \frac{24 \gamma_E -27+12 \ln2}{64 \pi }\langle b^4\rangle s^4 + \ldots
\end{equation}
This shows that the cumulants of $P(h)$ beyond the variance do not vanish (they depend on the {\it moments} of the source parameter distribution) and, conseqeuently,
$P(h)$ does not {\it uniformly} converge to a Gaussian (a $h^{-4}$ tail is always present for any finite $N_0\gg 1$).

We evaluate the asymptotic expansion for $P(h)$ at large (positive) $h$ using a Mellin transform argument, again. Equation \eqref{eqn:G Mellin small s} implies (bear in mind that $G(s)$ is even) that
\begin{equation}\label{eqn:approximation for P in terms of Q}
  P(h) \sim Q(h) \equiv \frac{1}{\pi} \int_{0}^{\infty}\mathrm{d}s\, \cos(hs)\exp\left(-\frac{3N_0\langle b^2\rangle}{8\pi}s^2 + \frac{N_0\langle b^3\rangle}{3\pi}s^3 + \frac{24 \gamma_E -27+12 \ln2}{64 \pi }N_0\langle b^4\rangle s^4\right) \;.
\end{equation}
Taking the Mellin transform of this equation gives
\begin{equation}
  \bar{Q}(\mu) = \int_{0}^{\infty}\mathrm{d}h \int_{0}^{\infty}\mathrm{d}s\, h^{\mu-1}\cos(hs)\exp\left(-\frac{3N_0\langle b^2\rangle}{8\pi}s^2 + \frac{N_0\langle b^3\rangle}{3\pi}s^3 + \frac{24 \gamma_E -27+12 \ln2}{64 \pi }N_0\langle b^4\rangle s^4\right).
\end{equation}
The $h$ integral is just
\begin{equation}
  \int_{0}^{\infty}\mathrm{d}h\, h^{\mu-1}\cos(hs) = s^{-\mu}\Gamma(\mu)\cos\left(\frac{\pi\mu}{2}\right)\;,
\end{equation}
when $0 < \Re \mu < 1$. The $s$ integral may be written as
\begin{equation}
  \int_{0}^{\infty} \mathrm{d}s\, s^{-\mu} \exp\left(-\frac{3N_0\langle b^2\rangle}{8\pi}s^2\right)\sum_{n=0}^{\infty}\frac{1}{n!}\left(\frac{N_0\langle b^3\rangle}{3\pi}s^3\right)^n\sum_{k=0}^{\infty}\frac{1}{k!}\left(\frac{24 \gamma_E -27+12 \ln2}{64 \pi }N_0\langle b^4\rangle s^4\right)^k\;.
\end{equation}
Each integral is of the form
\begin{equation}
  \int_{0}^{\infty}\mathrm{d}s\, s^{-\nu} \exp\left(-\frac{3N_0\langle b^2\rangle}{8\pi}s^2\right) = \frac{1}{2}\left(\frac{3N_0\langle b^2\rangle}{8\pi}\right)^{(\nu-1)/2}\Gamma\left(\frac{1}{2}-\frac{\nu}{2}\right)\;,
\end{equation}
where $\nu = \mu - 3n - 4k$. The Mellin transform inversion theorem implies (again) that for $0< c = \Re(\mu) < 1$
\begin{equation}\label{eqn:Mellin inversion Q}
  Q(h) = \frac{1}{2\pi \mathrm{i}}\int_{c-\mathrm{i}\infty}^{c+i\infty}\mathrm{d}\mu\, h^{-\mu}\bar{Q}(\mu)\;.
\end{equation}
The expression for $\bar{Q}$ has poles at $\mu = 0, -2, -4,\ldots$ due to $\Gamma(\mu)\cos(\mu \pi/2)$, as well as poles at $\nu = 1,3,5,\ldots$. The former set of poles does not affect the large $h$ expansion, as one needs to consider poles with $\Re(\mu) \geq 1$. The cosine cancels any $\nu$ pole unless $\mu$ is even, whence $n$ has to be odd for there to be a non-zero residue. As $\mu$ increases with $n$, the low values of $n$ contribute lower powers of $1/h$ to the expansion of $Q$ at $h \gg 1$. The lowest such $\mu$ for which there is a non-zero residue therefore dominates. It comes from $n = 1, k = 0$, and is given by $\mu = 4$ (the residues from $\mu = 1,2$ and $3$ are zero), which corresponds to $\nu = 1$. This yields
\begin{equation}
  Q(h) = -\left(-\frac{12}{2}\right)\frac{N_0\langle b^3\rangle}{3\pi}h^{-4}
\end{equation}
on using
\begin{equation}
  \text{Res}\!\left(\cos \left(\frac{\pi  z}{2}\right) \Gamma (z) \Gamma \left(\frac{1}{2}-\frac{1}{2} (-4 k-3 n+z)\right),\{z = 4\}\right) = -12\;.
\end{equation}
An additional minus sign comes from the fact that, when the line integral in the Mellin transform inversion formula (equation \eqref{eqn:Mellin inversion Q}) is shifted all the way to the right, it translates into clockwise contour integrals around the poles. Together with equation \eqref{eqn:approximation for P in terms of Q} this yields
\begin{equation}
  P(h) \sim \frac{2N_0\langle b(\tau = 0)^3 \rangle}{\pi h^4},
\end{equation}
which is precisely equation \eqref{eqn:h^-4 power law physical} (the $\tau = 0$ dependence is added as in Sec.~\S\ref{sec:physical case}).

It is clear that powers of $s$ higher than $s^3$ do not affect this scaling as they contribute only larger values of $\mu$ and, therefore, are subdominant.


\section{Numerical integration and parameters}
\label{appendix:integral}

As explained in Sec.~\S\ref{sec:physical case}, we evaluate the eight-dimensional integral, \cref{eqn:G}, as well as the various normalization constants numerically, before performing the inverse Fourier transform to obtain $P(h)$. For the integration, we use Monte-Carlo integration. In particular, we employ importance sampling, i.e. the VEGAS algorithm. Since the integrand is well-behaved, convergence can be reached very quickly, outperforming nested integration in this particular case. The integration in the $T$ variable is carried out logarithmically. For the final plots we used $10^8$ function evaluations to sample the integral. As a sanity check, we compared it to nested integration and obtained excellent agreement. We tabulate the values for $G(s)$, and interpolate logarithmically to perform the final, one-dimensional inverse Fourier transformation. At high values of $h$ the inverse Fourier transform can pick up modes from the numerical noise of the Monte-Carlo integration, but by increasing the number of function evaluations the value of $h$ where this starts to happen can be pushed to ever higher values.
For all integrations, we used the routines provided by the GSL \cite{gsl_manual}.

\begin{figure}
  \centering
  \includegraphics[width=0.45\textwidth]{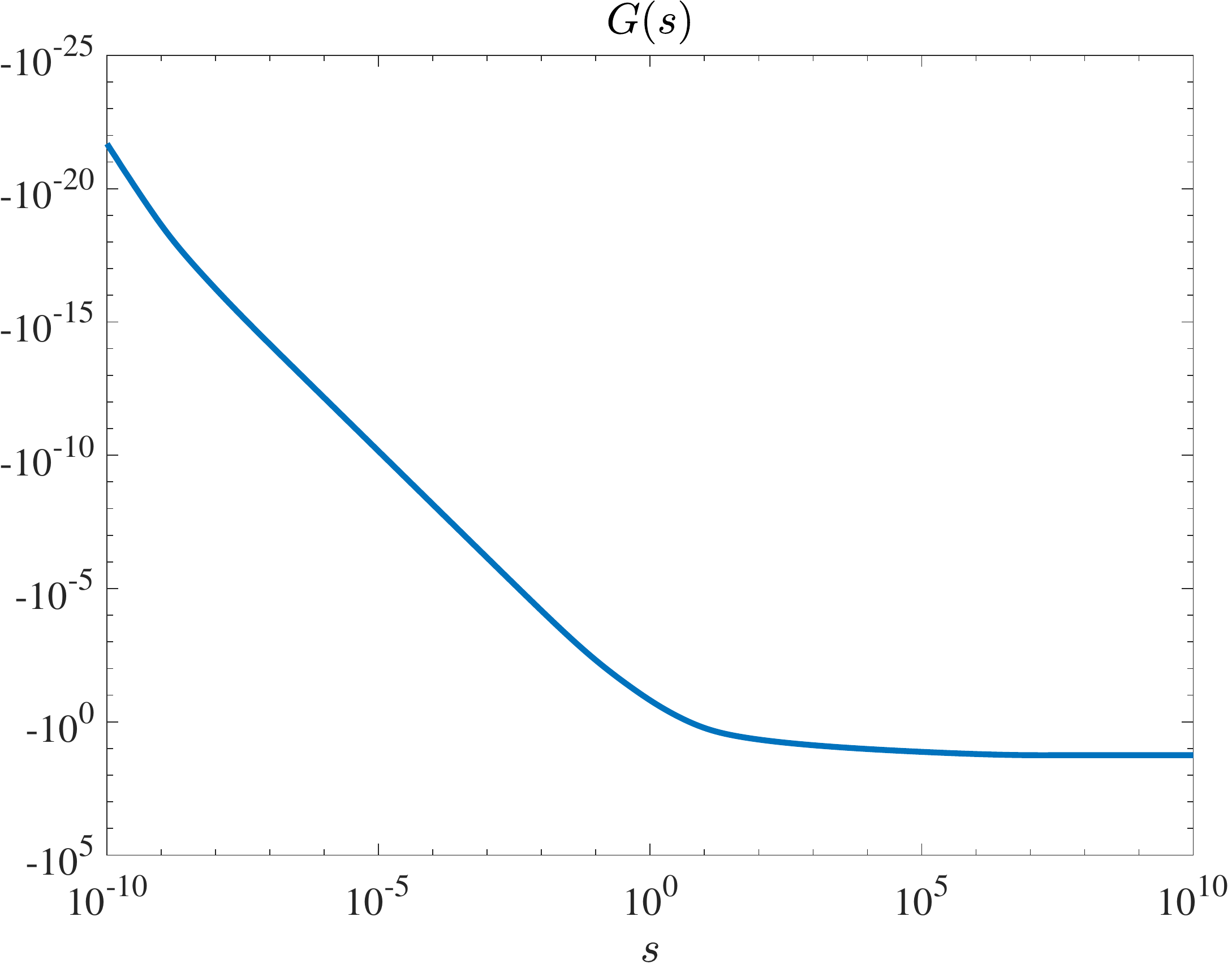}
  \caption{The value of $G(s)$ (equations \eqref{eqn:single source Fourier} and \eqref{eqn:G}) for the parameters specified in table \ref{table:constraints}.}
  \label{fig: G physical r0=0.01}
\end{figure}

All the parameters entering in the evaluation of the integral in equation \eqref{eqn:G} are summarized in \cref{table:constraints}.
A short distance cutoff $r_\text{min}=$0.01 Mpc was introduced only to avoid possible numerical issues. It does not affect our conclusions in any way.

\begin{table}
\centering
\begin{center}
\begin{tabular}{cccc}
Parameter $\ \ $  & Value  &  Units &  Description  \\ \hline\hline
$R$ &  $14$ & Gpc  & Co-moving size of the Universe.\\
$r_\mathrm{min}$ &  0.01 & Mpc  & UV cut-off on co-moving distance from detector.\\
$m_\mathrm{max}$ &  45 & $M_\odot$  & Maximum mass of binary member.\\
$m_\mathrm{min}$ &  8 & $M_\odot$  & Minimum mass of binary member.\\
$a_\mathrm{max}$ &  4000 & AU & Maximum initial semi-major axis of binary.\\
$a_\mathrm{min}$ &  0.014 & AU  & Minimum initial semi-major axis of binary.\\
$r_0$ & 10 & Gpc  & Attenuation length.\\
\end{tabular}
\caption{Summary of the integration boundaries, the corresponding units and additional parameters appearing in the numerically evaluated integral, \cref{eqn:G}.}
\label{table:constraints}
\end{center}
\end{table}

\section{Shape of Exponential Decline in \S \ref{sec:confusion background}}
\label{appendix:exponential asymptotic}

By introducing an effective cut-off in $G(s)$ due to bright sources, as described in \S \ref{sec:confusion background}, $G(s)$ is made into an analytic function, and the $\abs{s}^3$ irregular term is regularized. This implies that, by the Paley-Wiener theorem, its Fourier transform, $P_r(h)$, must decline faster than any power-law at large values of $h$, i.e., the power-law approximation ceases to hold for sufficiently large $h$. The purpose of this appendix is to find the asymptotic limit of $P_r(h)$ in the case where $G(s)$ is analytic.

The integral we need to approximate is given by equation
\eqref{eqn:stochastic gw probability density}, where $\psi(z)$ is an even
function of $z$, a task we perform using the steepest descents method \cite{Olver1974}. This method relies on
finding the saddle points of the exponent, and then deforming the integration
contour in the complex plane so that it coincides with the steepest descents
contour (on which the imaginary part of the exponent is constant) that passes
through the relevant saddle points. The remaining parts of the integral would
give exponentially small errors. In our case, the
exponent is $\zeta(z) = -\mathrm{i}hz + G(z)$, whence the saddle points are
those where $G'(z) = \mathrm{i}h$. The relevant steepest
descents contour coincides with the imaginary axis, for $G$ is even. Therefore, we set $z =
-\mathrm{i}y$ so that $\zeta = -hy + G(-\mathrm{i}y)$. Moreover, since $G(0)
= 0$ and $G$ is an entire function, then $z$ has to be large for $G'$ to be
as large as $\mathrm{i}h$. This justifies the replacement of
$G(-\mathrm{i}y)$ by its asymptotic expansion as $\abs{y} \to \infty$. In the
particular case at hand, it is of the form
\begin{equation}\label{eqn:asymptotic for G}
  -hy + c\exp(ay)y^{-b},
\end{equation}
where $a,b,c > 0$ are independent of $y$.

The reason for equation \eqref{eqn:asymptotic for G} is as follows: $\tilde{p}$, as given by equation \eqref{eqn:single source Fourier}, is $J_0(\textrm{const} \times s)$ (where the constant is not a function of $s$), so $G(\mathrm{i}y)$ is the expectation value of $J_0(\textrm{const} \times \mathrm{i}y) = I_0(\textrm{const} \times y)$. When $y$ is large (as is necessary for a saddle point), $I_0(Ay) \sim \exp(Ay)(2\pi Ay)^{-1/2}$ \cite{DLMF}. One needs to calculate the expectation value over $A$, which, in the case of Sec.~\S\ref{sec:physical case}, is a multi-dimensional integral. As $y$ is large, this integral is well-approximated by Laplace's method: one may change variables so that $A$ is one of the integration variables; the value of $A$ is maximized at the boundary, so Laplace's method implies that this integration contributes one negative power of $y$ to the result. Therefore $b = \frac{3}{2}$. Furthermore, $a$ is the value of $A$ at the maximum. $c$ is the value of the expected value, evaluated at $a$, and integrated over all other parameters.

The critical points of $\zeta(y) = -hy+ce^{ay}/y^{3/2}$ are the solutions of
\begin{equation}
  h \sim \frac{ace^{ay}}{y^{1.5}}\;,
\end{equation}
bearing in mind that $y$ is large and retaining only the dominant terms. This equation is solved by the Lambert W-function, which is approximated by $ay \sim \ln h$ \cite{DLMF}. Applying Laplace's method yet another time to the $y$ integration, the steepest descents method yields
\begin{equation}\label{eqn:exponential decay}
  P_r(h) \sim \frac{1}{\sqrt{2\pi ch}}\frac{(\ln h)^{3/4}}{a^{7/4}}\exp\!\left[-\frac{h\ln h}{a}+ \frac{a^{3/2}ch}{(\ln h)^{3/2}}\right]\;.
\end{equation}
In \cref{fig:asymptotics} we show the three regimes of $P_r(x)$ ($x=h/h_c$); that is, the exact calculation, the intermediate power-law regime and the exponential cut-off at very large $x$. The latter is described by \cref{eqn:exponential decay}. The exponential curve and the powerlaw $x^{-4}$ are connected smoothly at some reference value $x_0$ upon adjusting the free constants $a$ and $c$ in \cref{eqn:exponential decay}.

\end{document}